\documentclass{aastex63}

\usepackage{subfigure}
\usepackage{booktabs}
\usepackage{multirow}
\usepackage{hyperref}

\begin{document}
\title{Distributions and Physical Properties of Molecular Clouds in the G24 Region of the Milky Way}

\correspondingauthor{Xi Chen}
\email{chenxi@gzhu.edu.cn}

\author{Tian Yang}
\affil{Center for Astrophysics, Guangzhou University, Guangzhou 510006, People’s Republic of China}

\author{Xi Chen}
\affil{Center for Astrophysics, Guangzhou University, Guangzhou 510006, People’s Republic of China}

\author{Xiao-Yun Xu}
\affil{Center for Astrophysics, Guangzhou University, Guangzhou 510006, People’s Republic of China}

\author{Yang Yang}
\affil{Center for Astrophysics, Guangzhou University, Guangzhou 510006, People’s Republic of China}

\author{En Chen}
\affil{Center for Astrophysics, Guangzhou University, Guangzhou 510006, People’s Republic of China}

\author{Jun Li}
\affil{Center for Astrophysics, Guangzhou University, Guangzhou 510006, People’s Republic of China}

\author{Bing-Gang Ju}
\affil{Purple Mountain Observatory, Chinese Academy of Sciences, Nanjing 210008, People’s Republic of China}

\author{Deng-Rong Lu}
\affil{Purple Mountain Observatory, Chinese Academy of Sciences, Nanjing 210008, People’s Republic of China}

\begin{abstract}

We report the spatial distribution and physical characteristics of molecular clouds in the G24 region, which is located near the intersection of the Milky Way's Galactic bar with the Norma arm and the 3 kpc arm. Utilizing molecular line data from the Milky Way Imaging Scroll Painting (MWISP) project, including $^{12}$CO, $^{13}$CO, and C$^{18}$O, along with our own observations of HCO$^{+}$ line using the Purple Mountain Observatory (PMO) 13.7 m telescope, we have revealed the complex architecture of molecular clouds in the G24 region. Seven giant molecular clouds, each with a mass exceeding $10^4$ $M_\odot$ and a typical H$_2$ column density of $10^{21}$ cm$^{-2}$, have been identified through observations of CO and its isotopes. The conversion factor $X_{\text{CO}}$ for the G24 region is estimated to be 8.25 $\times$ 10$^{19}$ cm$^{-2}$ (K km s$^{-1}$)$^{-1}$, aligning with the typical values observed in other regions. Adopting the GaussClumps algorithm, we have identified a total of 257, 201, and 110 clumps in $^{12}$CO, $^{13}$CO and C$^{18}$O within G24 region, respectively. The derived physical properties (including effective radius, mass, and virial parameter) indicate that the majority of these clumps are gravitationally bound, with a subset possessing the potential to form massive stars. Examination of gas infall activities within these clumps further suggests ongoing massive star formation. The complex physical and kinematic environment, shaped by the G24 region's unique location within the Milky Way, has limited the clear detection of gas outflows.
\end{abstract}


\section{Introduction} \label{sec:1}

The formation of massive stars ($M \geqslant 8 M_{\sun}$) has a significant influence on the dynamics and chemical evolution of galaxies, and their energetic radiation and winds can affect the state of the surrounding gas, promoting or inhibiting new star formation \citep{2007ARA&A..45..481Z}. Currently, there is no unified consensus on the formation mechanisms of massive stars due to the multiple difficulties in studying their formation processes, including short lifetimes, long distances, sparse observational samples, and complex circumstellar environments \citep{2018ARA&A..56...41M}. Molecular clouds are the primary site of star formation and are composed mostly of cold gas (mainly hydrogen) and dust. In local areas of high density and low temperature, gas and dust begin to collapse under gravity, forming a dense core. As these cores collapse, they get hotter and hotter, eventually reaching the point where nuclear fusion is possible to form stars \citep{1987ARA&A..25...23S}. Studying the distribution and properties of molecular clouds can help us understand the conditions required for the formation of massive stars.


Although hydrogen (H$_2$) is the most prevalent molecule in molecular clouds, direct observations of it are challenging due to the absence of a permanent electric dipole moment in the hydrogen molecule, resulting in very weak rotational transitions in the millimeter and submillimeter bands \citep{2013ARA&A..51..207B}. Carbon monoxide (CO), the second most abundant molecule after H$_2$, is more easily detectable and serves as an excellent tracer for investigating molecular clouds and star formation processes \citep{1970ApJ...161L..43W}. The use of CO's different isotopes allows us to probe regions of varying densities: $^{12}$CO is ideal for mapping molecular cloud envelopes with densities around 10$^2$ cm$^{-3}$, $^{13}$CO is suitable for detecting regions of moderate density between 10$^2$ and 10$^3$ cm$^{-3}$, and C$^{18}$O effectively tracks dense molecular cores with densities up to 10$^4$ cm$^{-3}$ \citep{2017PASJ...69...78U}. Furthermore, HCO$^{+}$, which is abundant and shows an enhanced abundance in ionized regions, is one of the commonly optically thick lines \citep{2011A&A...527A..88V}. HCO$^{+}$ is more directly linked to high-density regions compared to $^{12}$CO and has been proposed as a more effective tracer for infalling gas \citep{2020RAA....20..115Y}. The synergistic use of multiple molecular tracers enables a more comprehensive understanding of the distribution of molecular gases and their dynamic processes.

\begin{figure}[ht]
    \centering
    \includegraphics[width=0.6\linewidth]{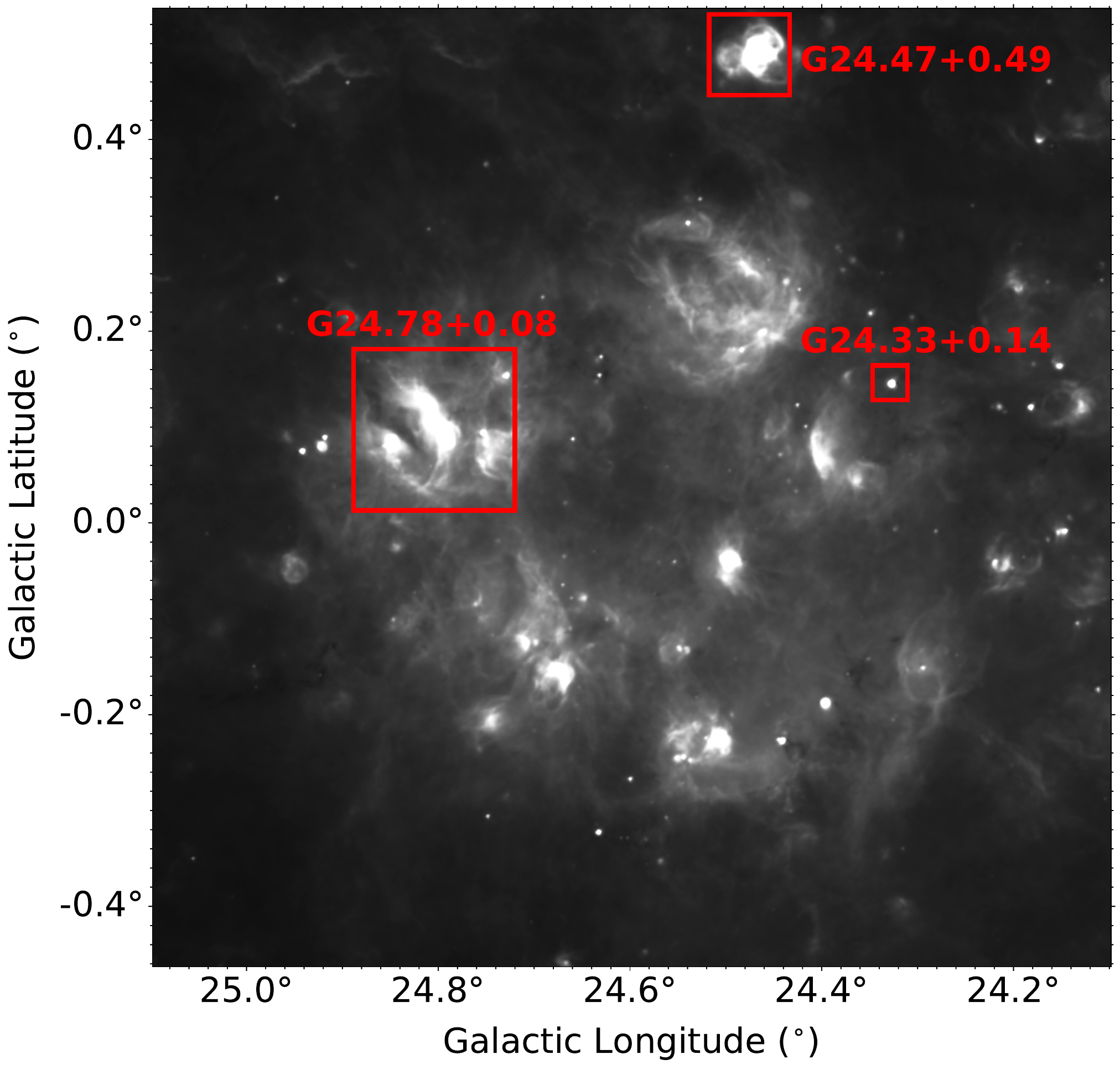}
    \caption{An infrared image of the Herschel 70 $\mu m$ band \citep{2010PASP..122..314M}, with central coordinates $l$ = 24.599$^{\circ}$ and $b$ = $+$0.036$^{\circ}$, and the image size of 1$^{\circ}$ $\times$ 1$^{\circ}$. The red boxes mark the locations of three MSFRs.}
    \label{fig:G24 infrared image}
\end{figure}

The G24 region encompasses several active massive star-forming regions (MSFRs) situated at the intersection of the Norma arm, the 3 kpc arm, and the near-end of the Galactic bar in the first quadrant of the Milky Way. The unique location of the G24 region, coupled with its rich gas and dust environment, offers invaluable insights into star formation processes and sheds light on the structure and evolution of the Milky Way. Figure \ref{fig:G24 infrared image} displays an infrared image of the region, which shows a ring-shaped distribution of several luminous molecular clouds, visually reminiscent of a cat's paw. Several MSFRs within the G24 region are under intensive investigation. G24.33$+$0.14 is acknowledged as a novel luminosity-bursting high-mass young stellar object (HMYSO), discovered through persistent monitoring of the 6.7 GHz methanol maser \citep{2023A&A...671A.135K,2023ApJ...951L..24L,2022PASJ...74.1234H,2022MNRAS.509.1681M,2019MNRAS.484.1590R}. G24.78$+$0.08 is encircled by multiple extended H{\sc ii} regions, driving the bipolar outflow phenomenon \citep{2018ApJ...866...20D,2011A&A...532A..91B,2007A&A...471L..13B,2006Natur.443..427B}. G24.47$+$0.49 serves to investigate the hierarchical triggering feature and its connection to multi-epoch star formation \citep{2024ApJ...970L..40S}. Despite fruitful research in the G24 region, particularly regarding star formation activities, the large-scale molecular cloud structure has rarely been examined. The Milky Way Imaging Scroll Painting (MWISP) project\footnote{\url{http://www.radioast.nsdc.cn/mwisp.php}} using the Purple Mountain Observatory (PMO) has now completed its CO observations in this region, and we will leverage these observations of $^{12}$CO, $^{13}$CO, and C$^{18}$O molecules to conduct a study on the distribution and physical properties of large-scale molecular clouds.

The structure of the paper is outlined as follows: Section \ref{sec:2} presents the CO MWISP project data and the HCO$^{+}$ molecular line observations obtained from the PMO 13.7 m telescope. Section \ref{sec:3} details the distributions and physical properties of the molecular clouds. In Section \ref{sec:4}, we delineate the CO clumps and explore their physical properties, as well as describe the infall and outflow phenomena within the G24 region. Conclusively, Section \ref{sec:5} provides a summary of our findings.

\section{Molecular line data} \label{sec:2}
\subsection{MWISP} \label{sec:2.1}

The observations of CO and its isotopic molecules in the G24 region were conducted as part of the MWISP project organized by the PMO. The observations utilized the 13.7 m telescope located in Delingha, Qinghai Province, China, at an altitude of 3200 meters \citep{2019ApJS..240....9S}. The survey simultaneously observed the J=$1-0$ transition of $^{12}$CO and its isotopes $^{13}$CO and C$^{18}$O. The first phase of the survey, covering the area with galactic longitudes $l$ = [-10$^{\circ}$, 250$^{\circ}$] and latitudes $b$ = [-5\fdg2, 5\fdg2] degrees, has ended in April 2021 after a decade-long observation period. The second phase commenced in September 2021, expanding the survey's coverage to latitudes $b$ = [-10$^{\circ}$, 10$^{\circ}$] degrees, with the anticipation that the survey will continue for another ten years.


The front-end uses a 3 $\times$ 3 beam sideband-separating Superconducting Spectroscopic Array Receiver (SSAR) \citep{2012ITTST...2..593S}, and the back-end comprises 18 Fast Fourier Transform Spectrometers (FFTS). The upper sideband was used to observe $^{12}$CO (115.271 GHz), while the lower sideband was used to record both $^{13}$CO (110.201 GHz) and C$^{18}$O (109.782 GHz). Each FFTS has a bandwidth of 1 GHz, with velocity coverage of 2602 km s$^{-1}$ at 115.271 GHz, 2722 km s$^{-1}$ at 110.201 GHz, and 2732 km s$^{-1}$ at 109.782 GHz. Additionally, each FFTS has 16384 channels, each of which has a width of 61 KHz, and a corresponding velocity resolution of 0.159 km s$^{-1}$ for $^{12}$CO, 0.166 km s$^{-1}$ for $^{13}$CO, and 0.167 km s$^{-1}$ for C$^{18}$O. The survey was performed using a position-switch On-The-Fly (OTF) mapping mode. The observation area was segmented into multiple 30$\arcsec$ $\times$ 30$\arcsec$ cells, each of which was scanned in at least two orthogonal directions along galactic longitude and galactic latitude to reduce the scanning effect.


The antenna temperature ($T_{\text{A}}$) is calibrated using the standard chopper-wheel calibration method. The main-beam temperature ($T_{\text{mb}}$) can be obtained by the formula $T_{\text{mb}} = T_{\text{A}}/\eta_{\text{mb}}$, where the main-beam efficiency ($\eta_{\text{mb}}$) is 45.9\% at 115 GHz and 51.1\% at 110 GHz. Typical system temperatures ($T_{\text{sys}}$) are about 250 K in $^{12}$CO and about 140 K in $^{13}$CO and C$^{18}$O. The typical rms noise level in each image cell is 0.49 K for $^{12}$CO and 0.25 K for $^{13}$CO and C$^{18}$O. The calibration accuracy of the spectral line intensity is less than 10\%. The half-power beamwidth (HPBW) is 49$\arcsec$ for the $^{12}$CO line and 51$\arcsec$ for the $^{13}$CO and C$^{18}$O lines, with a pointing accuracy of about 5$\arcsec$. Data reduction is done with CLASS software from the GILDAS package\footnote{\url{https://www.iram.fr/IRAMFR/GILDAS/}}.

\subsection{\texorpdfstring{OTF mapping observations for HCO$^{+}$ molecule line}{OTF mapping observations for HCO+ molecule line}} \label{subsec:2.2}

From May to June 2022, we used the PMO 13.7 m telescope to conduct OTF mapping observations of the G24 region for the HCO$^{+}$ (J=1-0; 89.2 GHz) transition. The centre coordinate is 18h36m01.59s, $-$07d23m39.0s (J2000) and are mapped along the right ascension and declination with a size of 1 square degree. The FFTS, also with 1 GHz bandwidth and 16384 channels, has a velocity coverage of 3364 km s$^{-1}$ at 89.2 GHz, corresponding to a velocity resolution of 0.205 km s$^{-1}$ for HCO$^{+}$. HCO$^{+}$ line observations have $\eta_{\text{mb}}$ is 62.4\% and HPBW is 60$\arcsec$. The observation time is more than 50 hours. Spectral line data is also regridded into 30$\arcsec$ $\times$ 30$\arcsec$ pixels. $T_{\text{sys}}$ are in the range of 250 - 300 K, with a typical noise level of 0.05 K.

\section{Results} \label{sec:3}
\subsection{Overall Distributions of CO Molecular Clouds} \label{subsec:3.1}

\begin{figure}[ht]
    \centering
    \subfigure{\includegraphics[width=0.55\textwidth]{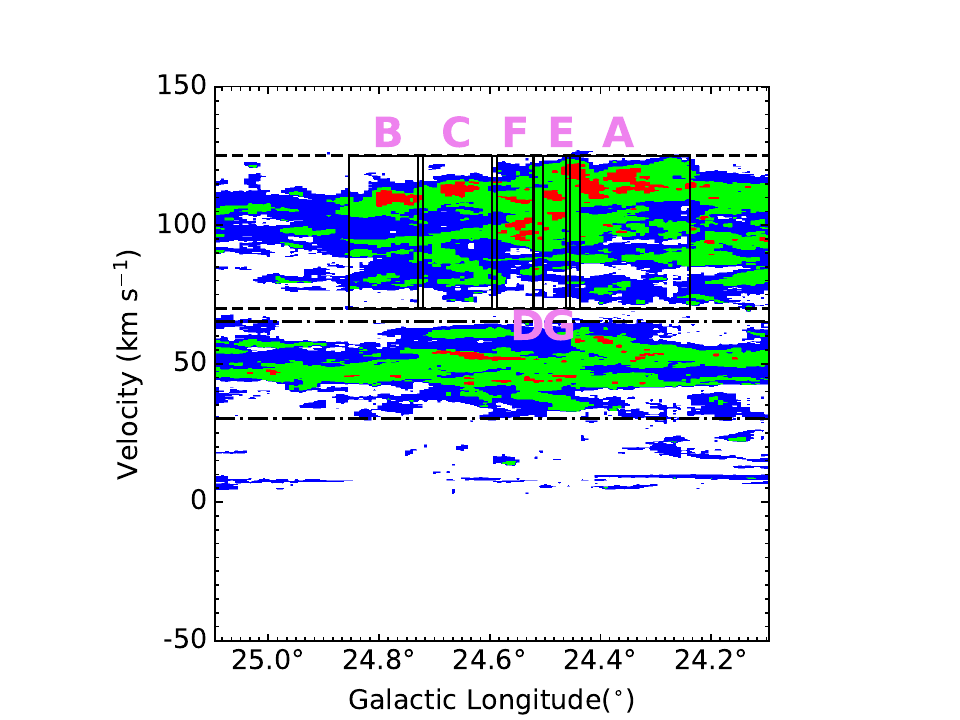}} 
    \hspace{-6.5em}
    \subfigure{\hspace{0.55\textwidth}} 
    
    \vspace{-1em}
    
    \subfigure{\includegraphics[width=0.55\textwidth]{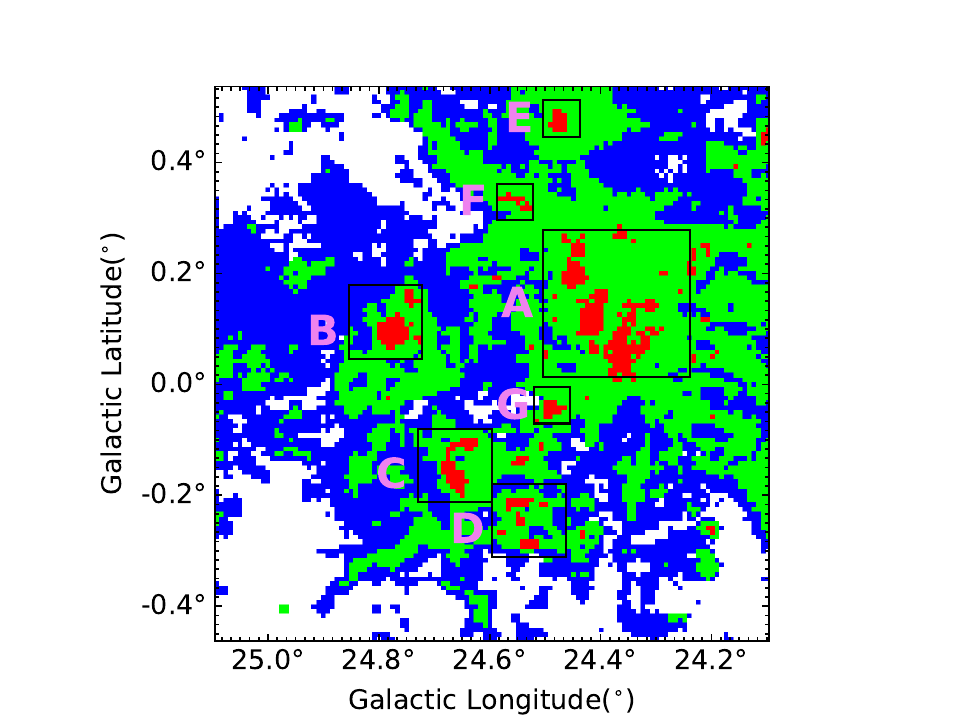}}
    \hspace{-6.5em}
    \subfigure{\includegraphics[width=0.55\textwidth]{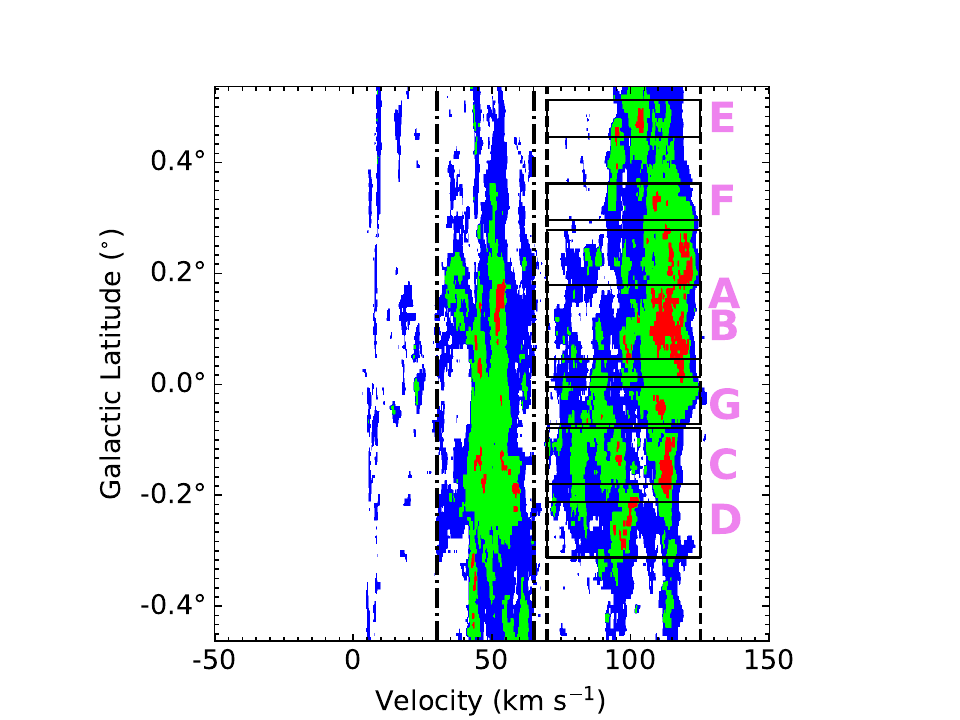}}
    \caption{Top left: longitude-velocity ($l$-$v$) map, integrating over Galactic latitude $b$ = $-$0.464$^{\circ}$ to 0.536$^{\circ}$. Bottom left: longitude-latitude ($l$-$b$) map, integrating over velocities $V_{lsr}$ = 70 to 125 km s$^{-1}$. Bottom right: velocity-latitude ($v$-$b$) map, integrating over Galactic longitude $l$ = 24.099$^{\circ}$ to 25.099$^{\circ}$. Blue, green, and red colors in all maps represent regions where $^{12}$CO, $^{13}$CO, and C$^{18}$O emission were detected, respectively. In the $l$-$v$ map and the $v$-$b$ map, there are two distinct velocity components, 30 - 65 km s$^{-1}$ and 70 - 125 km s$^{-1}$, the latter of which is the object of our study. 
    Based on the distribution characteristics of C$^{18}$O, we selected seven subregions for comparision and set the size of subregion A to 16$\arcmin$ $\times$ 16$\arcmin$, the size of subregions B, C, and D to 8$\arcmin$ $\times$ 8$\arcmin$, and the size of subregions E, F, and G to 4$\arcmin$ $\times$ 4$\arcmin$.}
    \label{fig:RBG_l-b-v}
\end{figure}

Figure \ref{fig:RBG_l-b-v} illustrates the $l$-$v$, $l$-$b$, and $v$-$b$ maps for the G24 region, obtained by integrating the latitude $b$ = $-$0.464$^{\circ}$ to 0.536$^{\circ}$, the longitude $l$ = 24.099$^{\circ}$ to 25.099$^{\circ}$, for the velocity $V_{lsr}$ = 70 to 125 km s$^{-1}$. In order to show the molecular cloud structure of this complex region more clearly, we set the weak voxel value to zero. Specifically, we retained only pixels with at least three consecutive channels with intensities greater than X $\cdot$ $\sigma$. For $^{12}$CO and $^{13}$CO, the value of X is set to 11, while for C$^{18}$O, which has a relatively weak intensity, X is 7. This process is called masking. 
In Figure \ref{fig:RBG_l-b-v} , the blue region shows the appearance of $^{12}$CO gas only, the green region indicates the detection of both $^{12}$CO and $^{13}$CO emission, and the red region represents the simultaneous presence of $^{12}$CO, $^{13}$CO, and C$^{18}$O. It should be noted that there are no cases in the G24 region where only $^{13}$CO is detected but not $^{12}$CO, or where only C$^{18}$O is detected but not $^{12}$CO and $^{13}$CO.


From the $l$-$v$ and $v$-$b$ maps, we observe two major and separate velocity components, located at 30 - 65 km s$^{-1}$ and 70 - 125 km s$^{-1}$, respectively. Of these, the 70 - 125 km s$^{-1}$ velocity component represents the region of interest for our study, which is located at the intersection of the 3 kpc arm and the bar, and also the starting point of the Norma arm. Based on the distribution characteristics of C$^{18}$O, we selected seven subregions with different sizes for a detailed study, as shown in the $l$-$b$ map of Figure \ref{fig:RBG_l-b-v}. 

\begin{figure}[ht]
    \centering
    \includegraphics[width=1\linewidth]{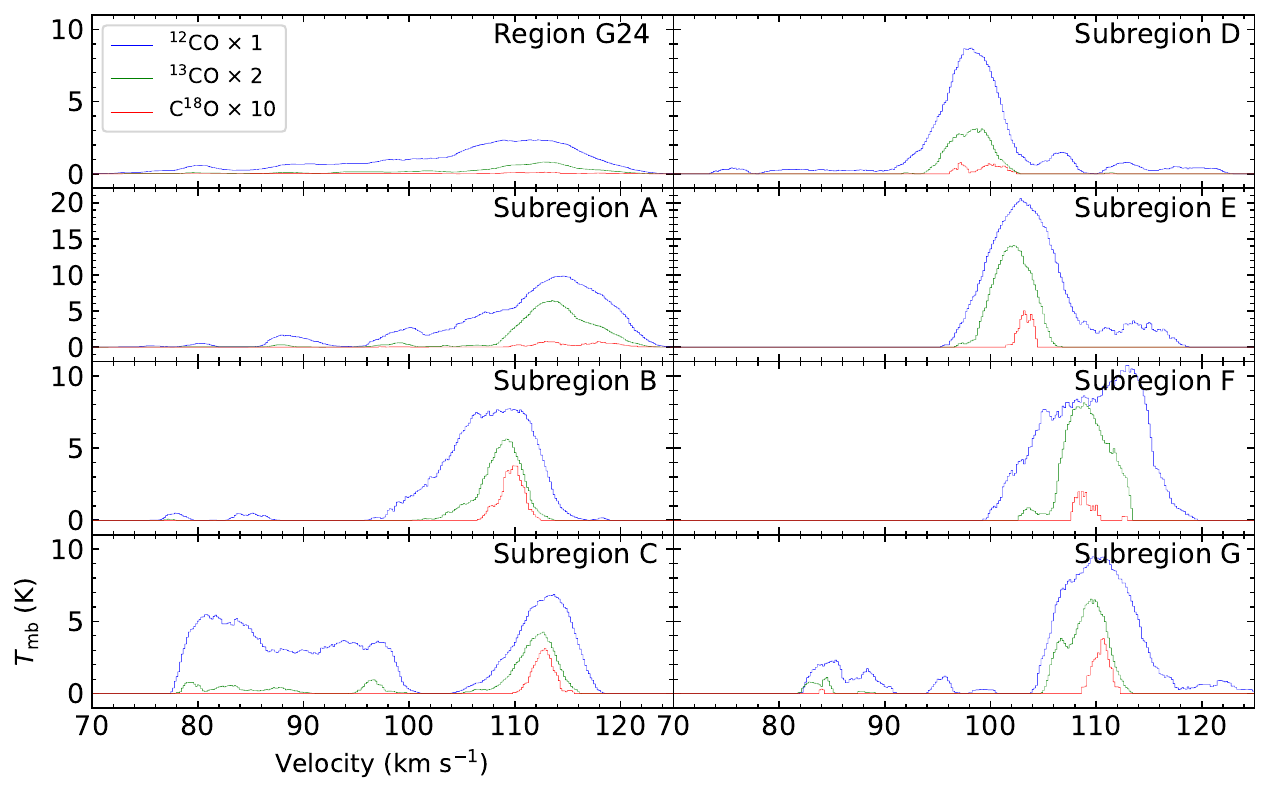}
    \caption{The average spectra of the G24 region and its seven subregions. Note that only pixels with intensities $\gtrsim$ X $\cdot$ $\sigma$ in at least three consecutive channels are averaged. X = 11 for $^{12}$CO and $^{13}$CO, X = 7 for C$^{18}$O.}
    \label{fig:AveragedSpectra}
\end{figure}

The average spectra of $^{12}$CO, $^{13}$CO, and C$^{18}$O for the G24 region and its seven subregions are shown in Figure \ref{fig:AveragedSpectra}. Table \ref{tab:observational properties} summarizes the line parameters for each subregion. The spectral line profile in the G24 region exhibits a wide range of velocities and multiple peaks, indicating that the molecular cloud composition in this region is extremely complex. The relatively small range of velocities detected in subregions E and F suggests a denser molecular cloud in these two subregions. Subregion E has the strongest peak emission intensity in $^{12}$CO, $^{13}$CO and C$^{18}$O, while its average integrated intensity in $^{12}$CO and $^{13}$CO is also the highest among all subregions. However, its average integrated intensity of C$^{18}$O is lower than that of subregions B and C. 

\begin{deluxetable}{ccccccccc}[ht]
\tablecaption{Observational properties of G24 region and seven subregions}
\tablehead{
\colhead{Region} & \colhead{Transitions} & \colhead{$V_{\text{min}}$} & \colhead{$V_{\text{max}}$} & \colhead{$V_{\text{pk}}$} & \colhead{$\Delta V$} & \colhead{$T_{\text{pk}}$} & \colhead{$I_{\text{tot}}$} & \colhead{$I_{\text{ave}}$} \\
\colhead{} & \colhead{} & \colhead{(km s$^{-1}$)} & \colhead{(km s$^{-1}$)} & \colhead{(km s$^{-1}$)} & \colhead{(km s$^{-1}$)} & \colhead{(K)} & \colhead{(K km s$^{-1}$ arcmin$^2$)} & \colhead{(K km s$^{-1}$)} \\
\colhead{(1)} & \colhead{(2)} & \colhead{(3)} & \colhead{(4)} & \colhead{(5)} & \colhead{(6)} & \colhead{(7)} & \colhead{(8)} & \colhead{(9)}}
\startdata
Region G24  & $^{12}$CO & 70.0  & 125.0 & 109.0 & 55.0 & 2.35  & 692155 & 253 \\
{}          & $^{13}$CO & 78.0  & 123.0 & 113.0 & 45.0 & 0.40  & 68605  & 53  \\
{}          & C$^{18}$O & 93.0  & 122.0 & 113.0 & 29.0 & 0.01  & 1188   & 11  \\
Subregion A & $^{12}$CO & 74.0  & 125.0 & 114.5 & 51.0 & 9.80  & 141784 & 554 \\
{}          & $^{13}$CO & 86.3  & 123.5 & 113.5 & 37.2 & 3.20  & 25394  & 104 \\
{}          & C$^{18}$O & 109.0 & 122.0 & 113.0 & 13.0 & 0.08  & 511    & 10  \\
Subregion B & $^{12}$CO & 76.5  & 119.0 & 109.5 & 42.5 & 7.80  & 20704  & 324 \\
{}          & $^{13}$CO & 100.0 & 114.0 & 109.2 & 14.0 & 2.85  & 3170   & 73  \\
{}          & C$^{18}$O & 106.5 & 112.5 & 109.9 & 6.0  & 0.38  & 245    & 19  \\
Subregion C & $^{12}$CO & 77.0  & 119.0 & 113.7 & 42.0 & 6.90  & 31518  & 492 \\
{}          & $^{13}$CO & 77.5  & 116.5 & 112.5 & 39.0 & 2.10  & 3099   & 62  \\
{}          & C$^{18}$O & 109.9 & 116.0 & 112.9 & 6.1  & 0.31  & 183    & 17  \\
Subregion D & $^{12}$CO & 73.0  & 123.0 & 98.0  & 50.0 & 8.70  & 17632  & 281 \\
{}          & $^{13}$CO & 91.5  & 103.0 & 99.3  & 11.5 & 1.55  & 1982   & 46  \\
{}          & C$^{18}$O & 96.0  & 103.0 & 97.2  & 7.0  & 0.08  & 65     & 9   \\
Subregion E & $^{12}$CO & 95.0  & 119.0 & 102.9 & 24.0 & 20.60 & 11071  & 692 \\
{}          & $^{13}$CO & 96.0  & 107.0 & 102.0 & 11.0 & 7.00  & 2083   & 130 \\
{}          & C$^{18}$O & 101.5 & 104.5 & 103.2 & 3.0  & 0.50  & 49     & 13  \\
Subregion F & $^{12}$CO & 99.0  & 120.0 & 113.0 & 21.0 & 10.80 & 7314   & 457 \\
{}          & $^{13}$CO & 102.5 & 113.7 & 109.0 & 11.2 & 4.10  & 1358   & 92  \\
{}          & C$^{18}$O & 107.5 & 113.0 & 109.0 & 5.5  & 0.20  & 22     & 7   \\
Subregion G & $^{12}$CO & 82.0  & 126.0 & 109.7 & 44.0 & 9.50  & 6026   & 389 \\
{}          & $^{13}$CO & 104.5 & 113.5 & 109.5 & 9.0  & 3.25  & 1003   & 67  \\
{}          & C$^{18}$O & 108.5 & 112.5 & 110.7 & 4.0  & 0.38  & 44     & 12 
\enddata
\tablecomments{Column (1): region name. Column (2): molecular transition. Columns (3)--(7): minimum velocity, maximum velocity, peak velocity, velocity range, and peak of the averaged spectra within the regions. Columns (8) and (9): total and average integrated intensities.}
\label{tab:observational properties}
\end{deluxetable}

\clearpage
\subsection{Physical Properties of CO Molecular Clouds} \label{subsec:3.2}

The G24 region contains multiple molecular clouds and the distances of these molecular clouds are different. The velocity range observed for G24.599$+$0.036 is between 70 and 125 km s$^{-1}$, which translates to a distance range of 5.26 [0.65] -- 7.16 [0.90] kpc when employing a rotational curve model to calculate the kinematic distances \citep{2019ApJ...885..131R}. To facilitate the analysis, we need to select a representative distance value for the G24 region. Currently, the most accurate method for determining distances within the Milky Way is the trigonometric parallax technique based on masers. Within the G24 region, we have identified two 6.7 GHz methanol maser sources with parallax measurement, G024.790$+$00.083 and G024.850$+$00.087, with corresponding parallax distance of 6.67 $\pm$ 0.71 kpc and 5.68 $\pm$ 0.52 kpc, respectively \citep{2019ApJ...885..131R}. Considering the high accuracy of the trigonometric parallax, we adopt the mean value of these two parallax distances, 6.18 ± 0.87 kpc, as the representative distance for the G24 region.

Assuming the CO molecular cloud is in local thermodynamic equilibrium (LTE) condition and treating $^{12}$CO as an optically thick line, the excitation temperature ($T_{\text{ex}}$) can be estimated by \citep{1998AJ....116..336N}:
\begin{equation}
    T_{\text{ex}} = \frac{5.53}{\ln(1+\frac{5.53}{T_{\text{pk,12}}+0.819})},
\end{equation}
where $T_{\text{pk,12}}$ is the peak main-beam temperature of $^{12}$CO.

The H$_{2}$ column density of the molecular cloud can be obtained by two different methods. The first method utilizes $^{13}$CO and C$^{18}$O molecules, which are assumed to have the same excitation temperature as $^{12}$CO under the assumption that they are in LTE condition (hereafter the LTE method). The second method is based on the $^{12}$CO molecule and estimates its H$_{2}$ column density by using the CO-to-H$_2$ conversion factor (hereafter the X-factor method).

For the LTE method, by the previous assumption we have obtained the excitation temperatures of $^{13}$CO and C$^{18}$O, and their optical depths ($\tau$) can be derived by \citep{2010ApJ...721..686P}:

\begin{equation}
    \tau_{13} = -\ln\bigg[1-\frac{T_{\text{pk,13}}}{5.29}\bigg(\frac{1}{\mathrm{e}^{5.29/T_{\text{ex}}}-1}-0.164\bigg)^{-1}\bigg],
\end{equation}
\begin{equation}
    \tau_{18} = -\ln\bigg[1-\frac{T_{\text{pk,18}}}{5.27}\bigg(\frac{1}{\mathrm{e}^{5.27/T_{\text{ex}}}-1}-0.166\bigg)^{-1}\bigg],
\end{equation}
where $T_{\text{pk,13}}$ and $T_{\text{pk,18}}$ are the peak main-beam temperatures of $^{13}$CO and C$^{18}$O, respectively. Then, the column densities ($N$) of $^{13}$CO and C$^{18}$O can be estimated as \citep{1997ApJ...476..781B}:

\begin{equation}
    N_{13} = 2.42 \times 10^{14} \cdot \frac{\tau_{13}}{1-\mathrm{e}^{-\tau_{13}}} \cdot \frac{1+0.88/T_{\text{ex}}}{1-\mathrm{e}^{-5.29/T_{\text{ex}}}} \cdot \int T_{\text{pk,13}}dv,
\end{equation}

\begin{equation}
    N_{18} = 2.54 \times 10^{14} \cdot \frac{\tau_{18}}{1-\mathrm{e}^{-\tau_{18}}} \cdot \frac{1+0.88/T_{\text{ex}}}{1-\mathrm{e}^{-5.27/T_{\text{ex}}}} \cdot \int T_{\text{pk,18}}dv.
\end{equation}


The isotopic ratios [$^{12}$C/$^{13}$C] = 6.21$R_{\text{GC}}$ + 18.71 \citep{2005ApJ...634.1126M} and [$^{16}$O/$^{18}$O] = 58.8$R_{\text{GC}}$ + 37.1 \citep{1994ARA&A..32..191W}, where $R_{\text{GC}}$ denotes the Galactocentric distance in units of kpc. The heliocentric distance of the G24 region is 6.18 kpc, which converts to a Galactocentric distance of 3.61 kpc according to the Galactic rotation curve model proposed by \citet{2019ApJ...885..131R}. The calculated isotope ratios are therefore [$^{12}$C/$^{13}$C] = 41 and [$^{16}$O/$^{18}$O] = 249. Finally, combining the abundance ratio [H$_2$/$^{12}$CO] = 1.1 $\times$ 10$^4$ \citep{1982ApJ...262..590F}, we can convert the column densities of $^{13}$CO and C$^{18}$O to the H$_2$ column density, respectively.


For the X-factor method, we start by defining $X_{\text{CO}}$ = $N_{\text{H$_2$,13}}$ / $I_{\text{CO}}$, where $N_{\text{H$_2$,13}}$ is the density of the H$_2$ column traced by $^{13}$CO obtained by the LTE method, and $I_{\text{CO}}$ is the integrated intensity of $^{12}$CO. By this method, we can derive the X-factor from the region where both $^{12}$CO and $^{13}$CO emission are detected, and then apply this X-factor to the entire region where $^{12}$CO emission is detected to calculate the H$_2$ column density of the $^{12}$CO molecule. A number of studies have shown that the X-factor varies with the environment \citep{2013ARA&A..51..207B,2015ARA&A..53..583H,2015ApJ...812....6B,2018ApJ...866...19B,2020ApJS..246....7S}. Therefore, we need to derive the value of the X-factor that is specific to the G24 region. 

\begin{figure}[ht]
    \centering
    \includegraphics[width=0.45\linewidth]{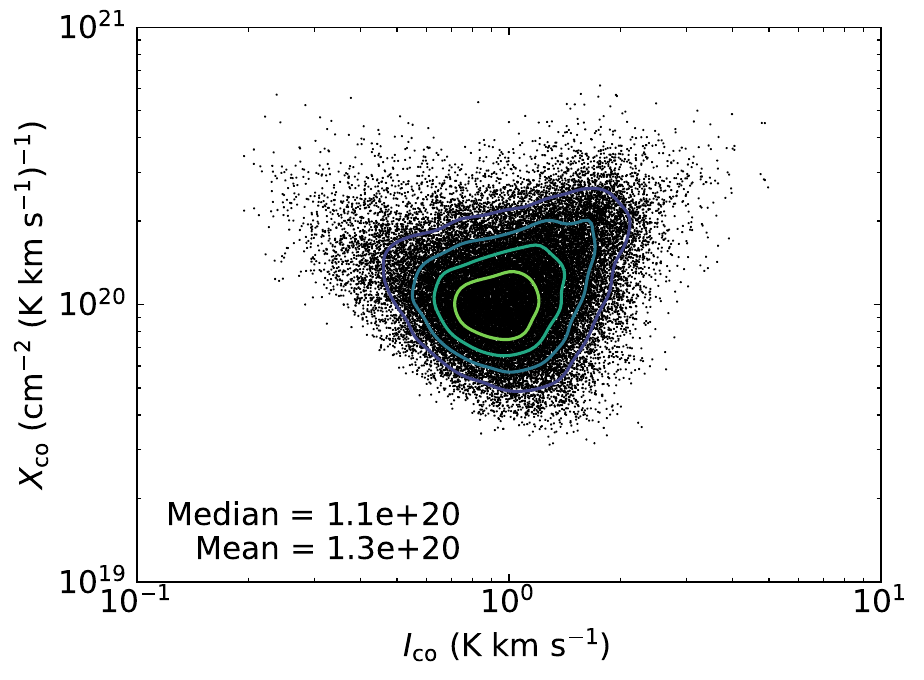}
    \includegraphics[width=0.45\linewidth]{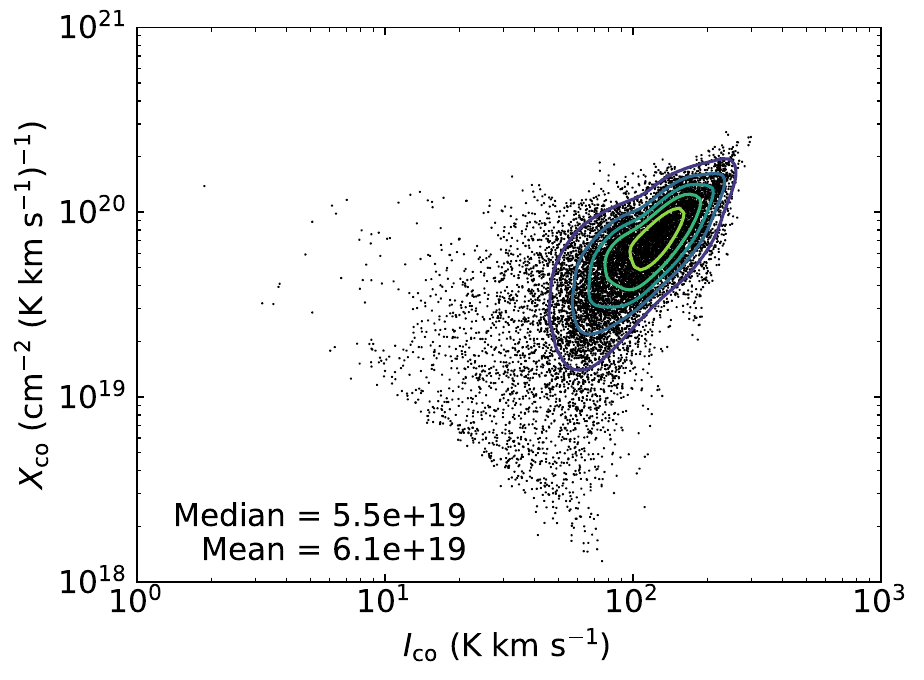}
    \caption{The relationship between $X_{\text{CO}}$ and $I_{\text{CO}}$. Left panel: each black dot represents a voxel, which detects both $^{12}$CO and $^{13}$CO.
    Right panel: each black dot represents a pixel that integrates over all velocity channels where $^{12}$CO emission is detected, regardless of whether $^{13}$CO emission is detected or not. The contours outline the surface densities of the black dots. The median and mean values of $X_{\text{CO}}$ are indicated in the lower left of each panel.}
    \label{fig:Xco}
\end{figure}

The relationship between $X_{\text{CO}}$ and $I_{\text{CO}}$ is presented in Figure \ref{fig:Xco}. 
Each black dot in the left panel of Figure \ref{fig:Xco} represents a voxel that contains both the $^{12}$CO and the $^{13}$CO emission. The panel exhibits an inverted V-shaped feature: at low $I_{\text{CO}}$ values, the relationship between $X_{\text{CO}}$ and $I_{\text{CO}}$ has a negative slope, while at high $I_{\text{CO}}$ values, it shows a positive slope. This feature is consistent with the findings of \citet{2020ApJS..246....7S} and \citet{2021ApJS..252...20L}. The voxel-derived $X_{\text{CO}}$ conversion factor can be considered as an upper limit of the X-factor. Each black dot in the right panel of Figure \ref{fig:Xco} represents a pixel that integrates over all velocity channels where $^{12}$CO emission is detected, regardless of whether $^{13}$CO emission is detected or not. \citet{2020ApJS..246....7S} argued that the pixel-derived $X_{\text{CO}}$ conversion factor represents a lower limit because pixels cover a wide range of velocities, resulting in many uncorrelated velocity components being integrated together, which contributes to some extent to the increase in $I_{\text{CO}}$, which in turn causes a decrease in $X_{\text{CO}}$. Thus, the actual $X_{\text{CO}}$ conversion factor may lie between the median values of 5.5 $\times$ 10$^{19}$ and 1.1 $\times$ 10$^{20}$. The average of this range, $X_{\text{CO}}$ = 8.25 $\times$ 10$^{19}$ cm$^{-2}$ (K km s$^{-1}$)$^{-1}$, was adopted for the G24 region in this paper. \citet{2024ApJ...971L...6S} used $X_{\text{CO}}$ = 1.0 $\pm$ 0.4 $\times$ 10$^{20}$ cm$^{-2}$ (K km s$^{-1}$)$^{-1}$ in their study of gas inflows in the Galactic region ($l$ = [1\fdg2, 19\fdg0]) using the MWISP CO data, which is very close to our G24 region, thereby reinforcing the reliability of our determined $X_{\text{CO}}$ value. \citet{2024PASJ...76..579K}, with the help of CO and its isotope data obtained from the FUGIN (FOREST Unbiased Galactic plane Imaging survey with the Nobeyama 45-m telescope) CO survey, revealed that the $X_{\text{CO}}$ values in the central molecular region (CMZ) range from (0.2 -- 1.3) $\times$ 10$^{20}$ cm$^{-2}$ (K km s$^{-1}$)$^{-1}$ and thus our $X_{\text{CO}}$ value falls within this range. In addition, based on the relationship found by \citet{1996PASJ...48..275A} for the variation of $X_{\text{CO}}$ with the galactocentric distance, we calculated $X_{\text{CO}}$ at a galactocentric distance of 3.61 kpc to be 1.0 $\times$ 10$^{20}$ cm$^{-2}$ (K km s$^{-1}$)$^{-1}$, which is in good agreement with our $X_{\text{CO}}$ \citep{2024MNRAS.527.9290K}.

Knowing the H$_2$ column density for each pixel, the total mass of the molecular gas can be obtained by integrating the entire region:
\begin{equation}
    M = 2 \mu m_{\text{H}} a^2 d^2 \sum N_{\text{H$_2$}},
\end{equation}
where $\mu$ = 1.36 is the average atomic weight of a hydrogen atom \citep{1983QJRAS..24..267H}, $m_{\text{H}}$ = 1.674 $\times$ 10$^{-24}$ g is the mass of a hydrogen atom, $a$ = 30{\arcsec} is the angular size of a pixel, and $d$ is the heliocentric distance in unit of pc. Note that the large $\sum$ symbol represents the summation symbol, which is used to sum the H$_2$ column densities of all pixels. The small $\Sigma$ symbol represents the surface density ($\Sigma$) of the molecular gas, and $\Sigma$ = $M$ / (a$^2$d$^2$$N_{\text{pixel}}$), where $N_{\text{pixel}}$ is the number of pixels. 

\begin{figure}[th!]
    \centering
    \includegraphics[width=0.9\linewidth]{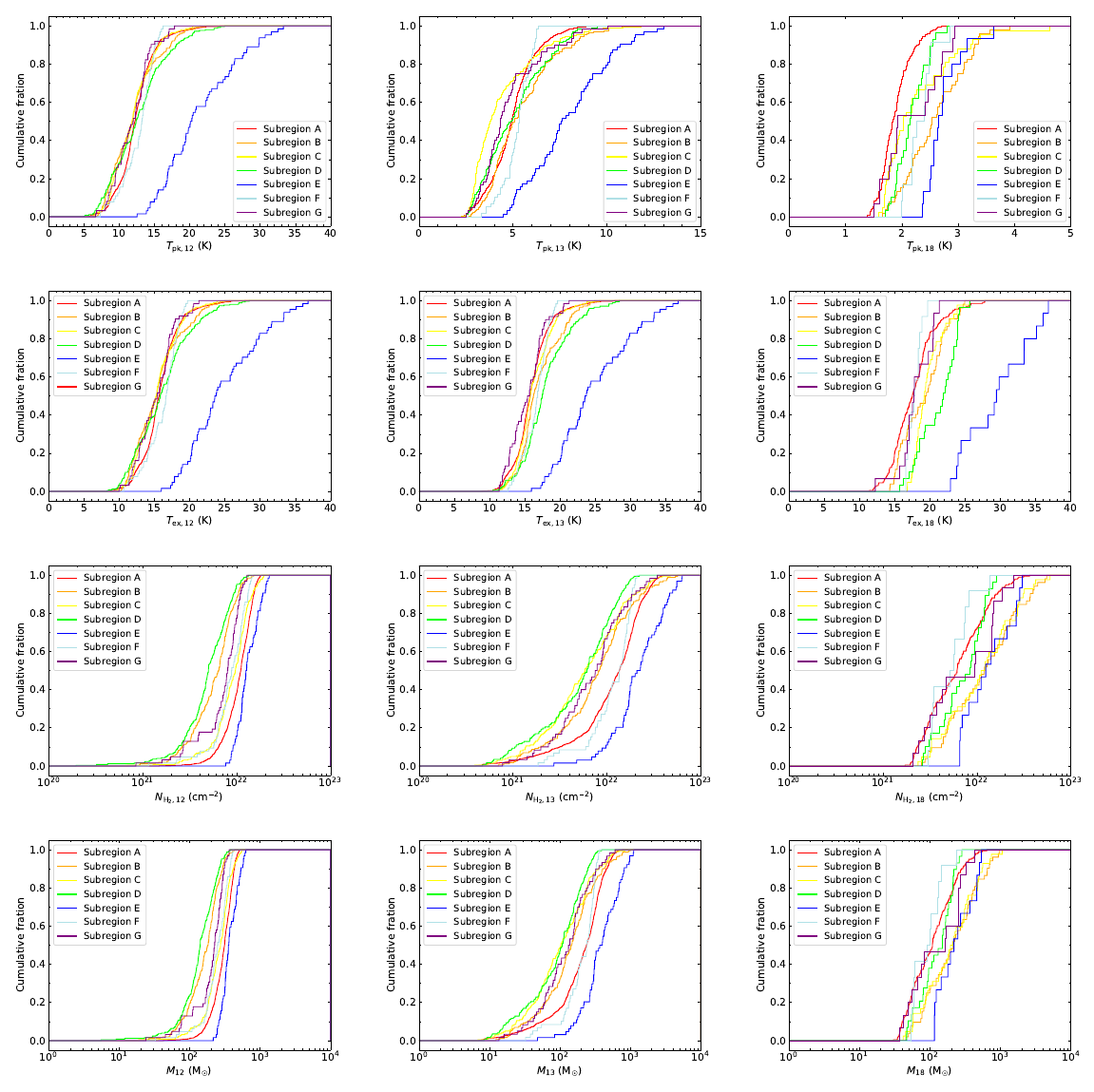}
    \caption{Cumulative distribution functions of the peak main-beam temperatures ($T_{\text{pk}}$), excitation temperature ($T_{\text{ex}}$), H$_2$ column density ($N_{\text{H$_2$}}$), and mass ($M$) of the $^{12}$CO, $^{13}$CO, and C$^{18}$O emission for the seven subregions. The first to fourth rows are $T_{\text{pk}}$, $T_{\text{ex}}$, $N_{\text{H$_2$}}$, and $M$, respectively, and the first to third columns denote the physical quantities of the $^{12}$CO, $^{13}$CO, and C$^{18}$O emission, respectively.}
    \label{fig:CDF_Tmb_Tex_NH2_M}
\end{figure}

Figure \ref{fig:CDF_Tmb_Tex_NH2_M} displays the distribution of peak main-beam temperature, excitation temperature, H$_2$ column density and mass for CO and its isotopes in the seven subregions. Table \ref{tab:physical properties} summarizes the various physical properties ($T_{\text{pk}}$, $T_{\text{ex}}$, $N_{\text{H$_2$}}$, $M$, $\Sigma$) of the G24 region and the seven subregions. As expected, as the gas becomes denser, from $^{12}$CO to $^{13}$CO, then to C$^{18}$O, the $T_{\text{pk}}$ gradually decrease, while the $T_{\text{ex}}$ gradually increase. Meanwhile, the $N_{\text{H$_2$}}$ obtained by $^{12}$CO, $^{13}$CO and C$^{18}$O, respectively, are essentially the same. The C$^{18}$O molecule mainly traces the dense gas structure inside the molecular cloud, while the $^{12}$CO molecule is utilized to probe the dilute gas at the outskirts of the molecular cloud. Therefore, the mass of the molecular cloud obtained from C$^{18}$O reflects the mass of the dense gas inside, while the mass of the molecular cloud obtained from $^{12}$CO represents the mass of the entire molecular cloud. Note that during the masking process in Section \ref{subsec:3.1}, many of the voxels with weaker emission were excluded due to the high threshold set, so the molecular cloud mass calculated here represents only a lower limit of the mass. In addition, there is significant variability in the physical properties between these subregions. Subregions A and D have low overall $T_{\text{pk}}$ distributions, showing relatively cooler characteristics. In contrast, in subregions B and C, the $T_{\text{pk}}$ values for C$^{18}$O are at their maximum. The most notable is subregion E, which has significantly higher values of $T_{\text{pk}}$, $T_{\text{ex}}$, $N_{\text{H$_2$}}$, $M$, and $\Sigma$ than the other subregions, reflecting the very active star formation activity in this subregion.

\setlength{\tabcolsep}{2pt}
\begin{deluxetable}{cccccccccccccccc}
\tablecaption{Physical properties of G24 region and seven subregions}
\tablehead{
\multirow{2}*{Region} & \multirow{2}*{Transitions} & \multicolumn{4}{c}{$T_{\text{pk}}$} & \multicolumn{4}{c}{$T_{\text{ex}}$} & \multicolumn{4}{c}{log($N_{\text{H$_2$}}$)} & \colhead{$M$} & \colhead{$\Sigma$}\\
\colhead{} & \colhead{} &\multicolumn{4}{c}{(K)} & \multicolumn{4}{c}{(K)} & \multicolumn{4}{c}{(cm$^{-2}$)} & \colhead{($M_{\sun}$)} & \colhead{($M_{\sun}$ pc$^{-2}$)} \\
\cmidrule(r){3-6} \cmidrule(r){7-10} \cmidrule(r){11-14}
\colhead{} & \colhead{} & \colhead{Min} & \colhead{Max} & \colhead{Mean} & \colhead{Median} & \colhead{Min} & \colhead{Max} & \colhead{Mean} & \colhead{Median} & \colhead{Min} & \colhead{Max} & \colhead{Mean} & \colhead{Median} & \colhead{} & \colhead{} \\
\colhead{(1)} & \colhead{(2)} & \colhead{(3)} & \colhead{(4)} & \colhead{(5)} & \colhead{(6)} & \colhead{(7)} & \colhead{(8)} & \colhead{(9)} & \colhead{(10)} & \colhead{(11)} & \colhead{(12)} & \colhead{(13)} & \colhead{(14)} & \colhead{(15)} & \colhead{(16)}}
\startdata
Region G24  & $^{12}$CO & 4.2  & 33.3 & 9.8  & 9.3  & 7.4  & 36.8 & 13.2 & 12.7 & 20.2 & 22.3 & 21.7 & 21.6 & 9.6 $\times$ 10$^5$ & 109 \\
            & $^{13}$CO & 2.1  & 13.0 & 4.4  & 4.1  & 8.5  & 36.8 & 15.2 & 14.8 & 20.6 & 22.8 & 21.9 & 21.6 & 6.5 $\times$ 10$^5$ & 154 \\
            & C$^{18}$O & 1.4  & 4.6  & 2.1  & 2.0  & 11.2 & 36.8 & 18.7 & 18.3 & 21.2 & 22.8 & 22.0 & 21.8 & 7.0 $\times$ 10$^4$ & 208 \\
Subregion A & $^{12}$CO & 5.4  & 24.3 & 12.4 & 12.2 & 8.7  & 27.8 & 15.8 & 15.6 & 21.2 & 22.3 & 22.0 & 22.0 & 2.0 $\times$ 10$^5$ & 238 \\
            & $^{13}$CO & 2.3  & 10.7 & 5.0  & 4.9  & 8.8  & 27.8 & 16.0 & 15.7 & 20.6 & 22.6 & 22.2 & 22.1 & 2.4 $\times$ 10$^5$ & 309 \\
            & C$^{18}$O & 1.4  & 2.7  & 1.9  & 1.8  & 11.4 & 27.8 & 17.7 & 17.5 & 21.2 & 22.6 & 21.9 & 21.8 & 2.8 $\times$ 10$^4$ & 173 \\
Subregion B & $^{12}$CO & 6.8  & 21.5 & 12.1 & 11.8 & 10.1 & 24.9 & 15.5 & 15.2 & 20.8 & 22.2 & 21.8 & 21.8 & 2.9 $\times$ 10$^4$ & 139 \\
            & $^{13}$CO & 2.7  & 11.3 & 5.5  & 5.1  & 11.3 & 24.9 & 17.0 & 16.3 & 20.8 & 22.8 & 22.0 & 21.9 & 3.4 $\times$ 10$^4$ & 242 \\
            & C$^{18}$O & 1.7  & 3.9  & 2.6  & 2.5  & 14.3 & 29.9 & 19.1 & 19.4 & 21.4 & 22.7 & 22.2 & 22.1 & 1.5 $\times$ 10$^4$ & 354 \\
Subregion C & $^{12}$CO & 6.0  & 22.1 & 11.9 & 11.8 & 9.3  & 25.6 & 15.4 & 15.2 & 20.8 & 22.3 & 22.0 & 22.0 & 4.4 $\times$ 10$^4$ & 211 \\
            & $^{13}$CO & 2.2  & 11.8 & 4.4  & 3.8  & 10.6 & 25.6 & 16.3 & 15.8 & 20.6 & 22.7 & 21.9 & 21.8 & 3.1 $\times$ 10$^4$ & 191 \\
            & C$^{18}$O & 1.6  & 4.6  & 2.3  & 2.0  & 16.8 & 25.6 & 19.7 & 19.2 & 21.4 & 22.8 & 22.2 & 22.1 & 1.1 $\times$ 10$^4$ & 334 \\
Subregion D & $^{12}$CO & 5.0  & 24.9 & 12.7 & 12.4 & 8.3  & 28.4 & 16.1 & 15.8 & 20.3 & 22.1 & 21.7 & 21.7 & 2.4 $\times$ 10$^4$ & 121 \\
            & $^{13}$CO & 2.5  & 9.5  & 5.1  & 4.9  & 10.5 & 28.4 & 17.9 & 17.5 & 20.7 & 22.3 & 21.8 & 21.8 & 2.1 $\times$ 10$^4$ & 151 \\
            & C$^{18}$O & 1.7  & 2.8  & 2.2  & 2.1  & 15.6 & 25.7 & 21.2 & 21.9 & 21.4 & 22.2 & 21.9 & 21.9 & 4.1 $\times$ 10$^3$ & 174 \\
Subregion E & $^{12}$CO & 12.5 & 33.3 & 21.4 & 20.2 & 16.0 & 36.8 & 24.9 & 23.7 & 21.9 & 22.3 & 22.1 & 22.1 & 1.5 $\times$ 10$^4$ & 297 \\
            & $^{13}$CO & 4.5  & 13.0 & 8.0  & 7.6  & 16.0 & 36.8 & 24.9 & 23.7 & 21.4 & 22.8 & 22.4 & 22.3 & 2.9 $\times$ 10$^4$ & 559 \\
            & C$^{18}$O & 2.4  & 3.6  & 2.7  & 2.7  & 22.9 & 36.8 & 29.6 & 29.4 & 21.8 & 22.5 & 22.2 & 22.1 & 4.1 $\times$ 10$^3$ & 334 \\
Subregion F & $^{12}$CO & 7.3  & 16.2 & 12.7 & 13.2 & 10.7 & 19.7 & 16.1 & 16.6 & 21.1 & 22.2 & 22.0 & 21.9 & 1.0 $\times$ 10$^4$ & 196 \\
            & $^{13}$CO & 3.3  & 6.4  & 5.2  & 5.3  & 12.6 & 19.7 & 16.6 & 16.8 & 21.3 & 22.3 & 22.1 & 22.1 & 1.3 $\times$ 10$^4$ & 275 \\
            & C$^{18}$O & 2.0  & 2.9  & 2.3  & 2.3  & 16.6 & 19.7 & 17.8 & 17.8 & 21.5 & 22.1 & 21.8 & 21.7 & 1.2 $\times$ 10$^3$ & 126 \\
Subregion G & $^{12}$CO & 6.6  & 17.8 & 11.8 & 12.0 & 10.0 & 21.3 & 15.2 & 15.4 & 20.9 & 22.1 & 21.9 & 21.9 & 8.4 $\times$ 10$^3$ & 167 \\
            & $^{13}$CO & 2.5  & 10.0 & 4.8  & 4.3  & 11.4 & 21.3 & 15.4 & 15.5 & 20.9 & 22.5 & 22.0 & 21.9 & 9.8 $\times$ 10$^3$ & 202 \\
            & C$^{18}$O & 1.5  & 2.9  & 2.2  & 1.9  & 12.2 & 21.3 & 17.9 & 17.7 & 21.3 & 22.4 & 22.0 & 22.0 & 2.5 $\times$ 10$^3$ & 209 
\enddata
\tablecomments{Column (1): region name. Column (2): CO transition. Columns (3) -- (14): minimum, maximum, mean, and median values of peak main-beam temperature ($T_{\text{pk}}$), excitation temperature ($T_{\text{ex}}$), and logarithm of H$_2$ column density (log($N_{\text{H$_2$}}$)). Column (15): total gas mass and surface density ($\Sigma$) derived from CO and its isotopes.}
\label{tab:physical properties}
\end{deluxetable}

\begin{figure}[ht]
    \centering
    \includegraphics[width=0.45\linewidth]{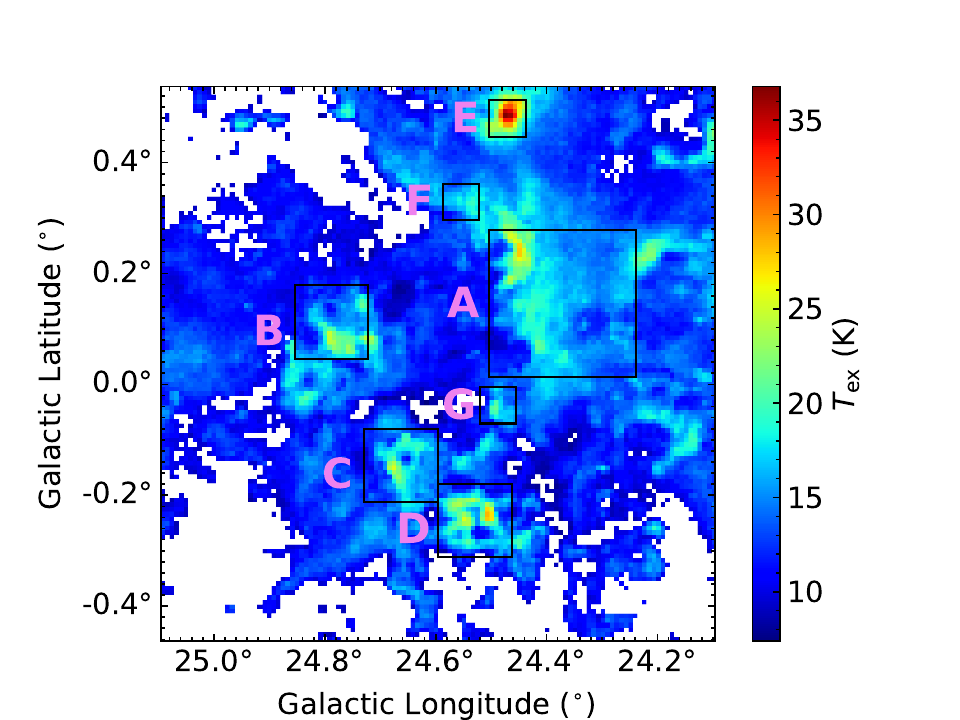}
    \hspace{-2em}
    \includegraphics[width=0.45\linewidth]{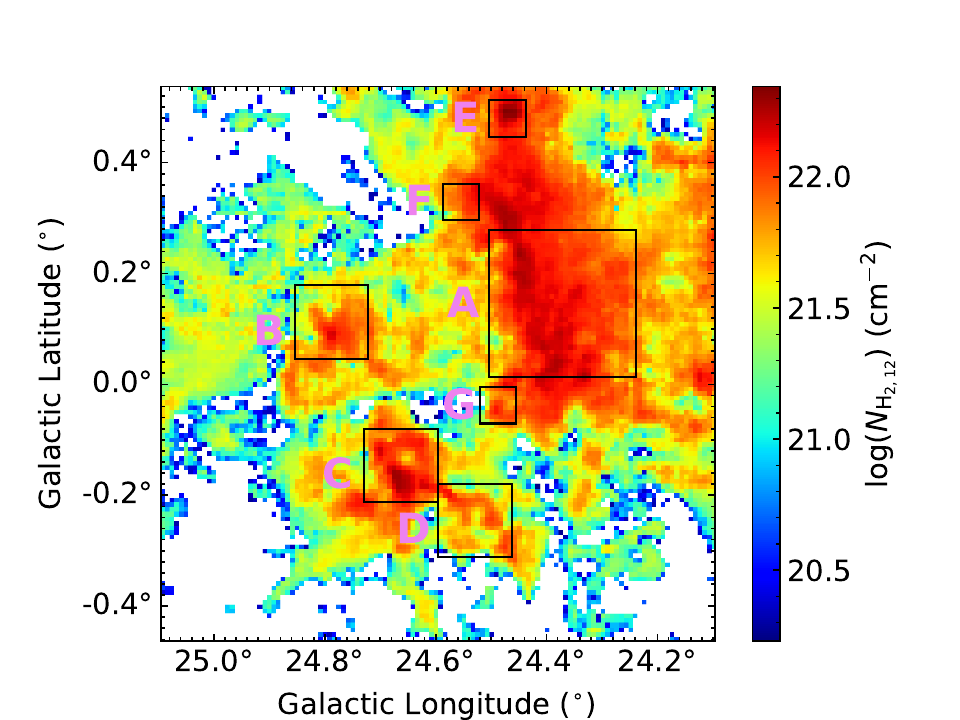}
    \includegraphics[width=0.45\linewidth]{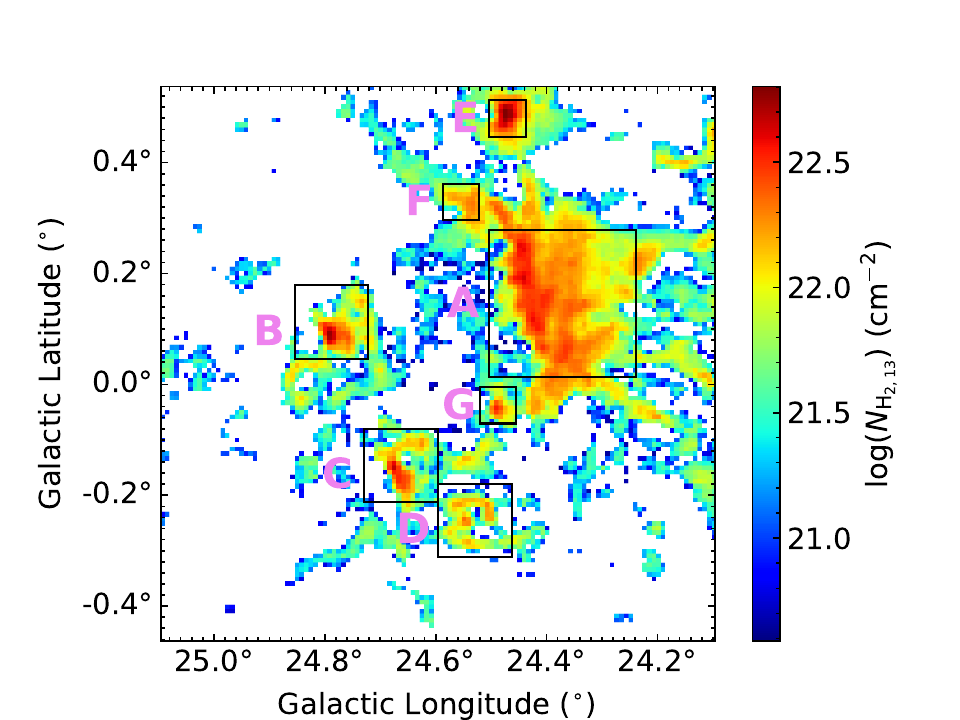}
    \hspace{-2em}
    \includegraphics[width=0.45\linewidth]{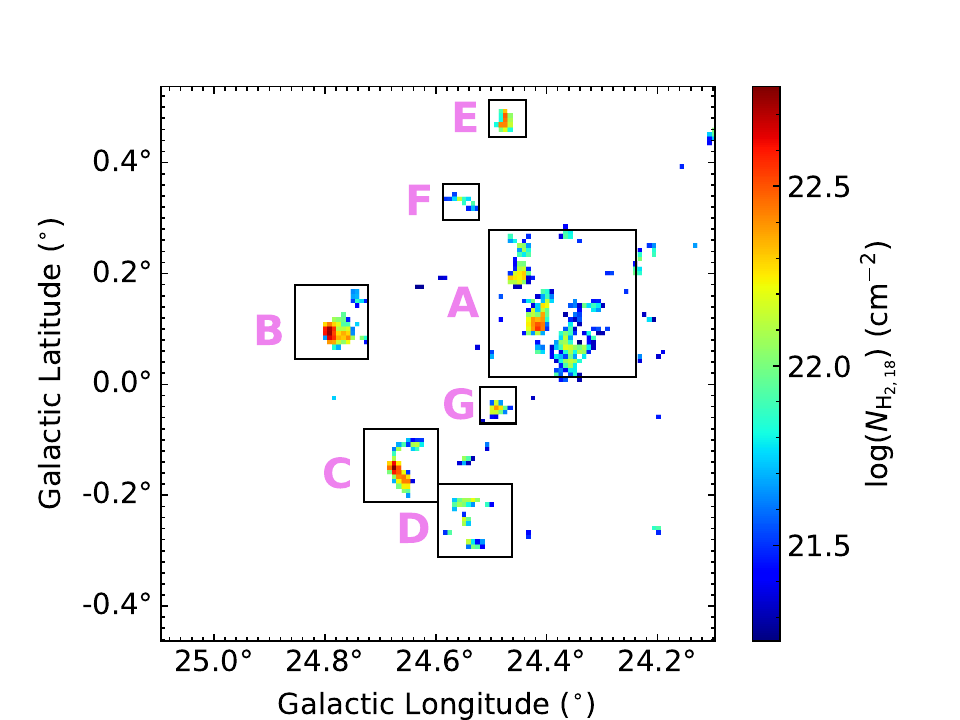}
    \caption{Upper left: Map of $T_{\text{ex}}$ distribution. Upper right: Map  of $N_{\text{H$_2,12$}}$ distribution. Lower left: Map  of $N_{\text{H$_2,13$}}$ distribution. Lower right: Map of $N_{\text{H$_2,18$}}$ distribution. Subregions A - G are labelled on each of the four maps.}
    \label{fig:map_of_Tex_N_distribution}
\end{figure}

Figure \ref{fig:map_of_Tex_N_distribution} shows the distribution of excitation temperatures and H$_2$ column densities traced by $^{12}$CO, $^{13}$CO, and C$^{18}$O, respectively. Majority of the regions have low excitation temperatures around 10 K, and only a few regions have excitation temperatures higher than 20 K, including subregion E with $T_{\text{ex}}$ as high as 36.8 K. In the H$_2$ column density distribution maps, the regions with higher densities show stripes, which appear to be filament structures. It is observable that regions of high excitation temperature generally coincide with areas of high density. 

\subsection{\texorpdfstring{HCO$^{+}$ molecule}{HCO+ molecule}} \label{subsec:3.3}

\begin{figure}[ht]
    \centering
    \includegraphics[width=0.85\linewidth]{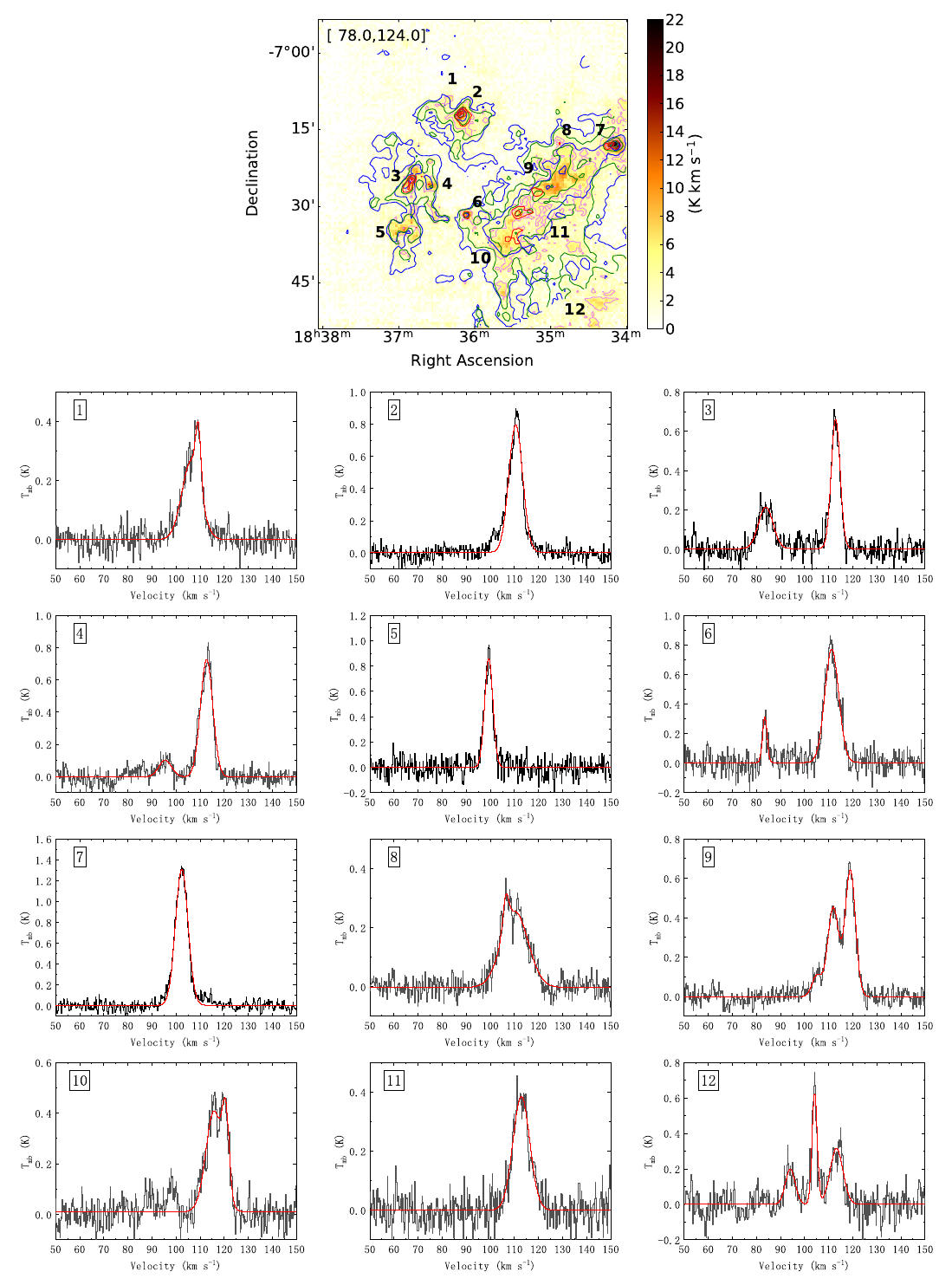}
    \caption{The HCO$^{+}$ (violet) emission region and the $^{12}$CO (blue), $^{13}$CO (green) and C$^{18}$O (red) emission regions from 78 -- 124 km s$^{-1}$, and the corresponding HCO$^{+}$ spectral line profiles. Upper panel: integrated intensity map for velocities from 78 -- 124 km s$^{-1}$ for coutours 20\%, 50\%, 80\%, with 12 regions with HCO$^{+}$ emission labeled 1 -- 12. Lower panel: HCO$^{+}$ spectral line profiles corresponding to the 12 regions.}
    \label{fig:mapping_78-124km/s_HCO+}
\end{figure}

Figure \ref{fig:mapping_78-124km/s_HCO+} presents the emission map and spectral line profiles of HCO$^{+}$ in the velocity range 78 - 124 km s$^{-1}$, and we have labeled 12 distinct HCO$^{+}$ emission regions. The spatial positions of these 12 regions of HCO$^{+}$ emission and their spectral parameters obtained by Gaussian fitting are given in Table \ref{tab:mapping_89.2GHzHCO+}. Some of the spectra contain two or three spectral components. The results show that the detected HCO$^{+}$ linewidths ranges from 2.3 to 12.8 km s$^{-1}$, the peak main-beam temperature ranges from 0.10 K to 1.31 K, and the integrated intensity ranges from 0.33 to 8.68 K km s$^{-1}$. 


The distributions of $^{12}$CO, $^{13}$CO, and C$^{18}$O are also superimposed on the HCO$^+$ data in Figure \ref{fig:mapping_78-124km/s_HCO+}. In general, there is a high degree of agreement between the HCO$^{+}$ and CO distributions, reflecting their similar forms of existence and evolution in the interstellar medium. However, there may still be differences in their radiative intensities and distribution details in some regions, which may be related to their respective chemical properties and physical environments. $^{12}$CO is widely distributed and intense due to its high abundance, while $^{13}$CO is similarly distributed but less intense, and C$^{18}$O is sparsely distributed and weakly intense due to its lower abundance and possible complex physicochemical effects. The distribution of HCO$^{+}$ lies between $^{13}$CO and C$^{18}$O, indicating that its distribution and radiative intensity in the interstellar medium are at an intermediate level.

\begin{deluxetable}{ccccccccccc}[ht]
\tablecaption{Mapping observations of HCO$^{+}$ molecule}
\tablehead{\colhead{Region} & \colhead{R.A. (J2000)} & \colhead{Dec. (J2000)} & \colhead{Line} & \colhead{$V_{\text{l}}$} & \colhead{$V_{\text{u}}$} & \colhead{$V_{\text{p}}$} & \colhead{Width} & \colhead{$T_{\text{p}}$} & \colhead{$I$} & \colhead{Rms} \\
\colhead{} & \colhead{(h m s)} & \colhead{($\circ$ $\prime$ $\prime\prime$)} & \colhead{} & \colhead{(km s$^{-1}$)} & \colhead{(km s$^{-1}$)} & \colhead{(km s$^{-1}$)} & \colhead{(km s$^{-1}$)} & \colhead{(K)} & \colhead{(K km s$^{-1}$)} & \colhead{(K)} \\
\colhead{(1)} & \colhead{(2)} & \colhead{(3)} & \colhead{(4)} & \colhead{(5)} & \colhead{(6)} & \colhead{(7)} & \colhead{(8)} & \colhead{(9)} & \colhead{(10)} & \colhead{(11)}}
\startdata
1  & 18 36 22.59 & $-$07 09 21.0 & 1 & 94.7  & 106.2 & 105.4 & 9.2  & 0.27 & 2.62 & 0.03 \\
   &             &               & 2 & 106.2 & 119.0 & 109.1 & 2.6  & 0.19 & 0.54 &      \\
2  & 18 36 08.26 & $-$07 11 01.0 & 1 & 97.5  & 120.0 & 110.4 & 6.9  & 0.79 & 5.82 & 0.03 \\
3  & 18 36 34.92 & $-$07 25 41.0 & 1 & 75.5  & 91.5  & 83.8  & 7.4  & 0.22 & 1.69 & 0.04 \\
   &             &               & 2 & 107.0 & 119.3 & 112.8 & 4.4  & 0.66 & 3.11 &      \\
4  & 18 36 48.26 & $-$07 25 41.0 & 1 & 88.7  & 103.3 & 95.4  & 6.3  & 0.10 & 0.69 & 0.04 \\
   &             &               & 2 & 103.3 & 121.4 & 112.6 & 6.1  & 0.73 & 4.69 &      \\
5  & 18 36 51.59 & $-$07 35 41.0 & 1 & 94.1  & 104.9 & 99.2  & 4.0  & 0.86 & 3.70 & 0.06 \\
6  & 18 36 08.26 & $-$07 31 51.0 & 1 & 80.7  & 86.6  & 83.4  & 2.3  & 0.31 & 0.76 & 0.06 \\
   &             &               & 2 & 102.0 & 120.7 & 111.2 & 6.9  & 0.77 & 5.65 &      \\
7  & 18 34 08.26 & $-$07 18 21.0 & 1 & 93.3  & 111.1 & 102.2 & 6.2  & 1.31 & 8.68 & 0.04 \\
8  & 18 34 44.92 & $-$07 18 21.0 & 1 & 96.0  & 109.1 & 106.3 & 2.9  & 0.11 & 0.33 & 0.03 \\
   &             &               & 2 & 109.1 & 125.6 & 110.0 & 12.8 & 0.26 & 3.49 &      \\
9  & 18 34 51.59 & $-$07 21 31.0 & 1 & 100.5 & 106.9 & 105.0 & 5.0  & 0.11 & 0.57 & 0.03 \\
   &             &               & 2 & 106.9 & 115.6 & 111.8 & 5.4  & 0.45 & 2.59 &      \\
   &             &               & 3 & 115.6 & 125.4 & 118.9 & 5.0  & 0.64 & 3.43 &      \\
10 & 18 35 34.92 & $-$07 37 11.0 & 1 & 106.1 & 117.1 & 115.6 & 8.3  & 0.25 & 3.58 & 0.04 \\
   &             &               & 2 & 117.1 & 125.8 & 120.6 & 3.4  & 0.41 & 1.06 &      \\
11 & 18 35 08.26 & $-$07 35 41.0 & 1 & 102.6 & 123.1 & 112.8 & 8.3  & 0.38 & 3.40 & 0.05 \\
12 & 18 34 21.59 & $-$07 49 31.0 & 1 & 88.4  & 100.0 & 93.9  & 4.9  & 0.20 & 1.03 & 0.06 \\
   &             &               & 2 & 100.0 & 107.5 & 104.0 & 2.6  & 0.62 & 1.71 &      \\
   &             &               & 3 & 107.5 & 120.3 & 113.1 & 6.3  & 0.31 & 2.12 &     
\enddata
\tablecomments{Column (1): Region number. Columns (2) and (3): positions in equatorial coordinates. Column (4): label for the different velocity components. Columns (5) -- (9): lower velocity limit, upper velocity limit, peak velocity,line width, and peak of the averaged HCO$^+$ line spectra within the region. Column (10): integrated intensity. Column (11): spetral rms noise.}
\label{tab:mapping_89.2GHzHCO+}
\end{deluxetable}

\clearpage
\section{Discussion} \label{sec:4}
\subsection{CO Clumps and Their Properties} \label{subsec:4.1}
\subsubsection{Identification of CO Clumps} \label{subsec:4.1.1}

Clumps usually refer to regions of high density in molecular clouds where the gravitational collapse of the gas occurs inside, eventually forming stars. There may be multiple clumps in a molecular cloud, each with the potential to form stars independently. 


The algorithm for identifying clumps used in this paper is GaussClumps built into CLASS software. The GaussClumps algorithm was developed by \citet{1990ApJ...356..513S} to identify and analyze clump structures in molecular clouds. The basic concept of the algorithm is to consider the observed molecular cloud data as a superposition of multiple Gaussian-shaped clumps and to identify and extract these clumps by fitting a Gaussian function. Specifically, the algorithm performs Gaussian fitting sequentially from the brightest peak position based on the magnitude of the intensity value. After each fit, the algorithm subtracts the fitted clumps from the original data and subsequently continues to find new peaks to fit in the remaining data until a preset threshold is reached. The algorithm allows for overlap between clumps, that is, a voxel can be assigned to more than one clump at the same time, which makes the algorithm more reliable in identifying molecular cloud clumps with complex structures.


Given the environmental complexity of the G24 region, we chose a higher threshold. This may miss some small mass clumps, but it won't affect our search for massive clumps capable of forming massive stars. 

\begin{figure}[ht]
    \centering
    \includegraphics[width=0.45\linewidth]{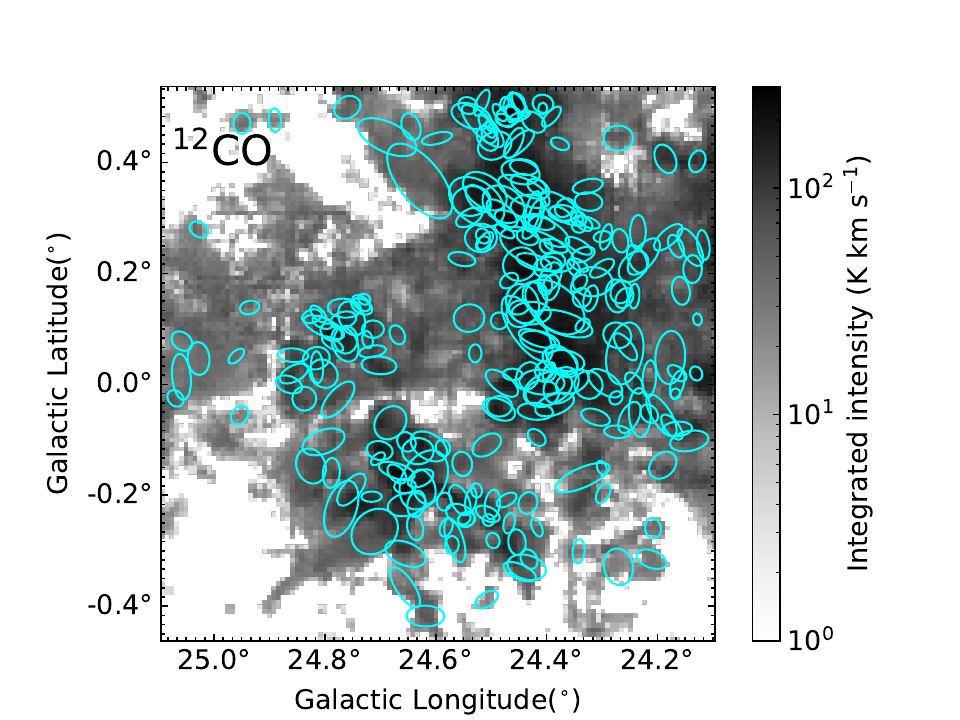}\\
    \includegraphics[width=0.45\linewidth]{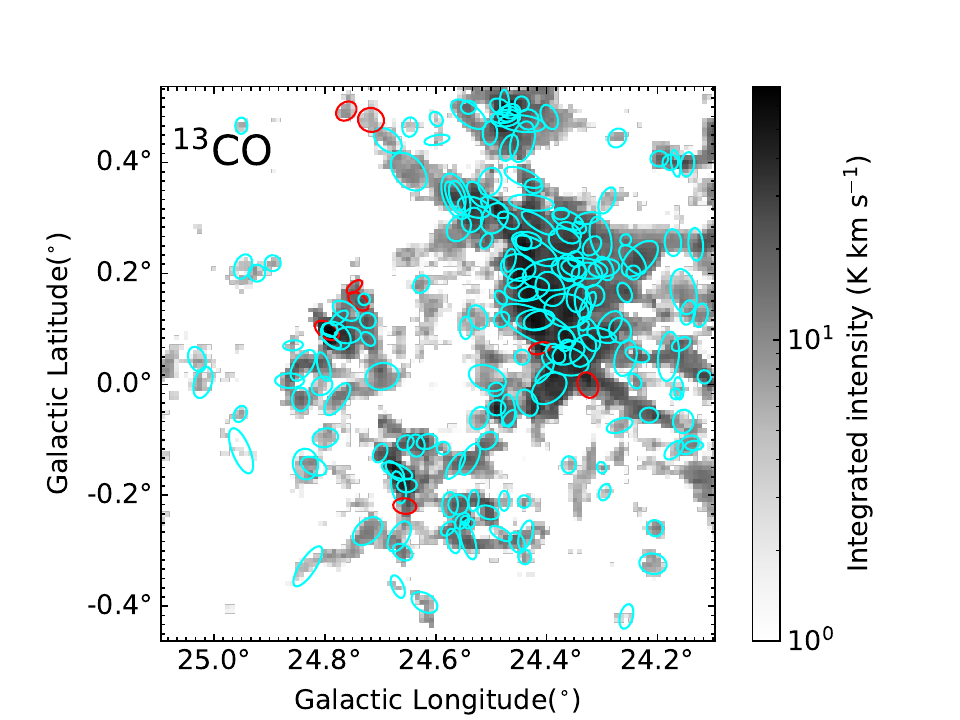}
    \includegraphics[width=0.45\linewidth]{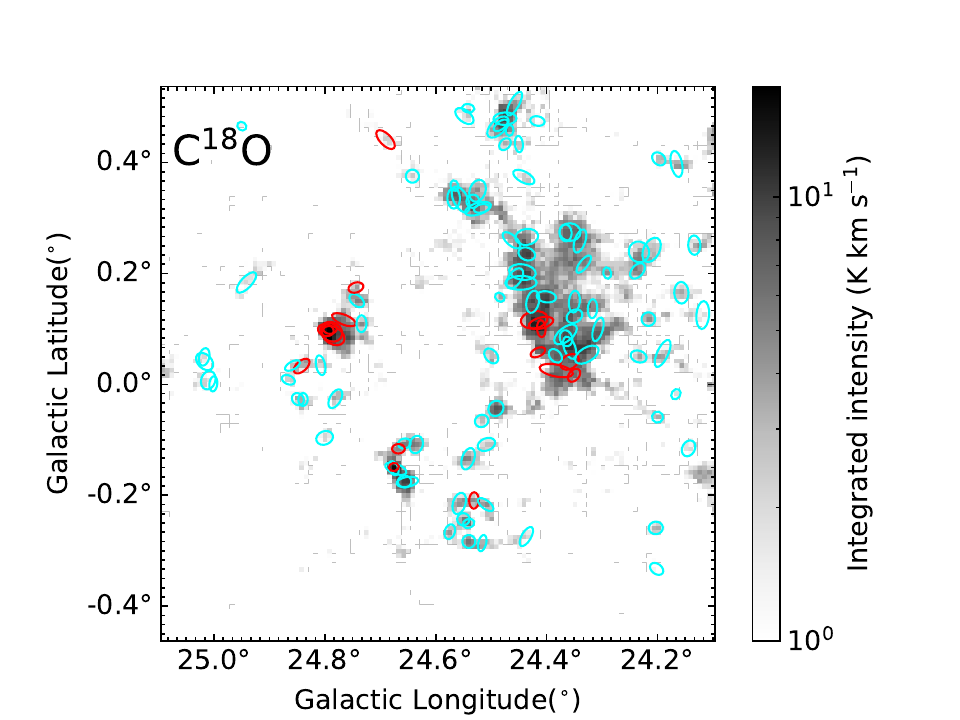}
    \caption{Spatial distributions of $^{12}$CO (top), $^{13}$CO (left), and C$^{18}$O (right) clumps. The background are the integrated intensity maps of the corresponding CO and its isotopes emission. The ellipses denote the positions of the clumps, where the cyan ellipses do not have the potential to form massive stars and the red ellipses indicate potential formation sites for massive stars, as detailed in Section \ref{subsec:4.1.5}.}
    \label{fig:Distribution_of_clumps}
\end{figure}

Notably, the clump result obtained using the GaussClumps algorithm is only preliminary and needs to be further screened to provide reliable data for subsequent clump analysis. The criteria for screening clumps are as follows: (1) clumps with sizes smaller than one beam are eliminated to ensure that the identified clumps are real within the resolution of the data used; (2) clumps with a ratio of the major axis to the minor axis greater than 3 are eliminated. This is because molecular clouds with a ratio of the major to minor axis greater than 3 are usually regarded as filamentary structure, which does not match the morphology of the clumps of focus in this study. After screening, the final numbers of $^{12}$CO, $^{13}$CO, and C$^{18}$O clumps were 257, 201 and 110, respectively. Figure \ref{fig:Distribution_of_clumps} depicts the spatial distributions of these clumps. Tables \ref{appendix:the_parameters_of_12CO_clumps}, \ref{appendix:the_parameters_of_13CO_clumps}, and \ref{appendix:the_parameters_of_C18O_clumps} list the measured parameters of the $^{12}$CO, $^{13}$CO, and C$^{18}$O clumps, respectively.

\subsubsection{Calculation of Physical Properties of CO Clumps} \label{subsec:4.1.2}
The excitation temperature ($T_{\text{ex}}$), optical depth ($\tau$) and column density ($N$) of the clumps are calculated by the same method as the molecular clouds, as detailed in Section \ref{subsec:3.2}. The difference is that in the column density calculation, the integrated intensity adopts $I = T_{\text{pk}} \sigma_v \sqrt{2 \pi} = \frac{1}{2} T_{\text{pk}} \Delta v \sqrt{\frac{\pi}{\ln2}}$, where $\sigma_v$ represents the velocity dispersion, $\Delta v$ denotes the linewidth, and are related as $\Delta v = \sqrt{8 \ln2} \sigma_v \approx 2.355 \sigma_v$.

The effective radius ($R_{\text{eff}}$) of a clump can be calculated by the following formula:
\begin{equation}
    R_{\text{eff}} = \frac{1}{2} d \sqrt{\theta_{\text{a}} \theta_{\text{b}}-\theta^2_{\text{beam}}},
\end{equation}
where $d$ is the source distance in unit of pc, $\theta_{\text{a}}$ and $\theta_{\text{b}}$ denote the major and minor axis of the Gaussian clump in unit of radians, respectively, and $\theta_{\text{beam}}$ is the beam size in unit of radians. The mass of the molecular cloud clump ($M_{\text{clump}}$) can be obtained by integrating over a three-dimensional Gaussian sphere in position-position-velocity space. The following is a simplified form:

\begin{equation}
    M_{\text{clump}} = 5.925 \times 10^{-31} \theta_{\text{a}} \theta_{\text{b}} d^2 N_{\text{H$_2$}} M_{\sun},
\end{equation}
where $N_{\text{H$_2$}}$ is the H$_2$ column density of the pixel corresponding to the center of the clumps, and the units of $\theta_{\text{a}}$ and $\theta_{\text{b}}$ are arcseconds.

The virial mass of the clump ($M_{\text{vir}}$) can be calculated by \citet{2016A&A...588A.104G}:
\begin{equation}
    M_{\text{vir}} = 5 \frac{\sigma^2_v R_{\text{eff}}}{G} \approx 209 \bigg(\frac{R_{\text{eff}}}{\text{pc}}\bigg) \bigg(\frac{\Delta v}{\text{km s$^{-1}$}}\bigg)^2 M_{\sun},
\end{equation}
where $G$ is the gravitational constant. The virial parameter ($\alpha_{\text{vir}}$) is defined as:
\begin{equation}
    \alpha_{\text{vir}} = \frac{M_{\text{vir}}}{M_{\text{clump}}},
\end{equation}
to characterize the gravitational state of the clump.

By treating the molecular cloud clumps as homogeneous spheres of effective radius $R_{\text{eff}}$, the surface density ($\Sigma$) and the number density of molecular hydrogen ($n_{\text{H$_2$}}$) can be calculated by the following equations, respectively:

\begin{equation}
    \Sigma = \frac{M_{\text{clump}}}{\pi R^2_{\text{eff}}},
\end{equation}

\begin{equation}
    n_{\text{H$_2$}} = \frac{M_{\text{clump}}}{\frac{4}{3} \pi R^3_{\text{eff}} 2\mu m_{\text{H}}},
\end{equation}
where $\mu$ = 1.36 is the average atomic weight of a hydrogen atom \citep{1983QJRAS..24..267H}, $m_{\text{H}}$ = 1.674 $\times$ 10$^{-24}$ g is the mass of a hydrogen atom.

The thermal linewidth ($\Delta v_{\text{th}}$) and the non-thermal linewidth ($\Delta v_{\text{nth}}$) can be calculated:

\begin{equation}
    \Delta v_{\text{th}} = \sqrt{8 \ln2 \frac{k T_{\text{k}}}{\mu_{\text{CO}} m_{\text{H}}}},
\end{equation}

\begin{equation}
    \Delta v_{\text{nth}} = \sqrt{\Delta v^2 - \Delta v^2_{\text{th}}},
\end{equation}
where $k$ is the Boltzmann constant, $T_{\text{k}}$ is the kinematic temperature of the gas, $T_{\text{k}}$ $\approx$ $T_{\text{ex}}$ under the LTE assumption and $\mu_{\text{CO}}$ is the molecular weight (28 for $^{12}$CO, 29 for $^{13}$CO, and 30 for C$^{18}$O). The physical parameters of $^{12}$CO, $^{13}$CO and C$^{18}$O clumps are listed in Tables \ref{appendix:the_parameters_of_12CO_clumps}, \ref{appendix:the_parameters_of_13CO_clumps} and \ref{appendix:the_parameters_of_C18O_clumps}, respectively.

\subsubsection{Statistics of Physical Properties of CO Clumps} \label{subsec:4.1.3}

\begin{figure}[ht]
    \centering
    \includegraphics[width=0.9\linewidth]{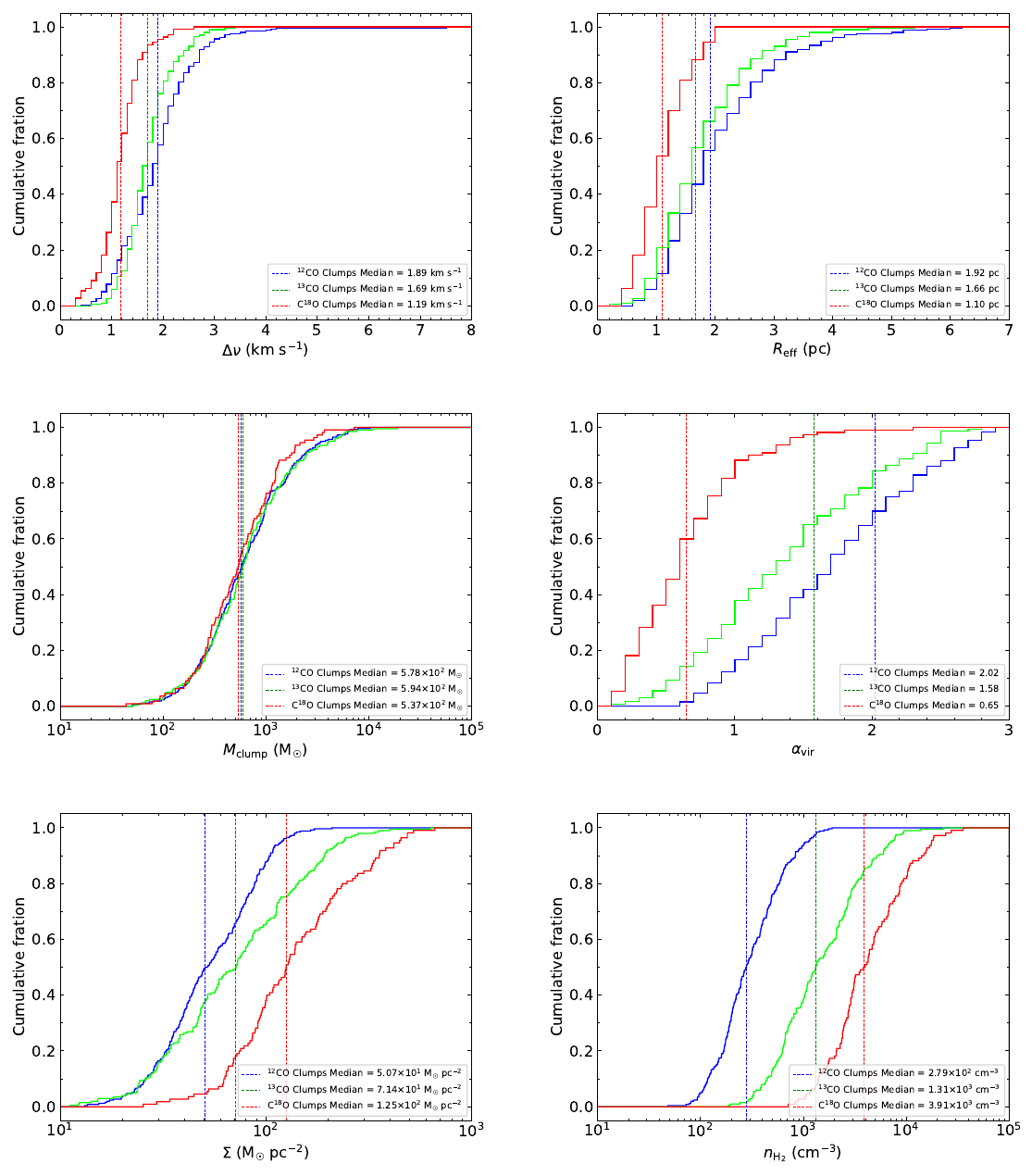}
    \caption{Cumulative distribution functions of physical parameters for $^{12}$CO, $^{13}$CO, and C$^{18}$O clumps, including linewidth ($\Delta v$), effective radius ($R_{\text{eff}}$), mass of clumps ($M_{\text{clump}}$), virial parameter ($\alpha_{\text{vir}}$), surface density ($\Sigma$), and number density ($n_{\text{H$_2$}}$). The median values of the physical parameters for the different CO isotopes are denoted by dashed lines and marked in the lower right corner of each panel.}
    \label{fig:CDF_clumps}
\end{figure}

Figure \ref{fig:CDF_clumps} illustrates the distributions of various physical parameters of the three CO clumps. From the top-left panel it can be seen that the clumps in the G24 region have a wide range of linewidth distributions, with the maximum linewidth of the $^{12}$CO clump as high as 7.51 km s$^{-1}$, and the median linewidths of the $^{12}$CO, $^{13}$CO, and C$^{18}$O clumps being 1.89, 1.69, and 1.19 km s$^{-1}$, respectively. In contrast, in the MWISP G220 region, the median velocity dispersion $\sigma_v$ of the $^{12}$CO, $^{13}$CO, and C$^{18}$O clumps is 0.4, 0.3, and 0.3 km s$^{-1}$, respectively, which corresponds to the linewidth of 0.94, 0.71, and 0.71 km s$^{-1}$, respectively \citep{2023ApJS..268....1D}. This suggests that the clump linewidths are larger in the G24 region, possibly due to the fact that the G24 region is closer to the Galactic center, has multiple active MSFRs, or complex gas motions. In addition, \citet{2014MNRAS.443.1555U} performed an extensive statistical analysis on massive star-forming (MSF) clumps using data from the ATLASGAL (The APEX Telescope Large Area Survey of the Galaxy) survey. They discovered that the median linewidth of MSF clumps is 2.2 km s$^{-1}$, which exceeds the linewidth of CO clumps in the G24 region. This discrepancy could be attributed to the typically more complex dynamical environments in which MSF clumps reside, such as molecular cloud collisions, compression, or gravitational collapse within their interiors. These processes can initiate stronger turbulent motions, resulting in increased non-thermal broadening of spectral lines. Additionally, feedback mechanisms following the formation of massive stars, including stellar winds and radiation, can disturb the surrounding gas, further augmenting the linewidth.


The top-right panel of Figure \ref{fig:CDF_clumps} shows that the $^{12}$CO, $^{13}$CO, and C$^{18}$O clumps have median effective radius of 1.92, 1.66, and 1.10 pc, with $^{12}$CO being the largest and C$^{18}$O the smallest, as expected. In the G220 region, the median effective radii of the $^{12}$CO, $^{13}$CO, and C$^{18}$O molecular clouds are measured at 0.9, 0.6, and 0.6 parsecs, respectively. These dimensions are considerably smaller than those observed in the G24 region. This discrepancy may be due to the G220 region's location in the outer reaches of the Milky Way, where interstellar material is relatively less dense \citep{2023ApJS..268....1D}. In addition, the ATLASGAL MSF clumps have a median effective radius of 0.95 pc, which is smaller than that of CO clumps because CO clumps are much larger-scale gas clouds containing multiple dense cores and filamentary structures, while MSF clumps tend to be these dense cores or the intersection of filamentary structures \citep{2014MNRAS.443.1555U}. 
Most of the Hi-GAL dense sources belong to clumps (with sizes between 0.2 and 3 pc) with a peak radius of about 0.5 pc \citep{2017MNRAS.471..100E,2021MNRAS.504.2742E}, and the reason for the small size of Hi-GAL clumps compared to CO clumps may be that the Hi-GAL survey probes dust clumps in star-forming regions, which are usually smaller because they are in the active phase of star formation. However, the MWISP observations cover a much wider region of the molecular cloud, and the clumps in these regions are likely to be larger because they encompass the entire molecular cloud structure.


The masses of the $^{12}$CO, $^{13}$CO, and C$^{18}$O clumps depicted in the middle-left panel of Figure \ref{fig:CDF_clumps} are primarily distributed between 10$^2$ and 10$^4$ $M_{\sun}$. The median masses of these three CO clumps are almost equal, between 500 and 600 $M_{\sun}$, which are much larger than the CO clouds in the G220 region \citep{2023ApJS..268....1D}. Meanwhile, the median mass of the ATLASGAL MSF clumps, with a value of about 1000 $M_{\sun}$, is nearly twice that of the CO clumps in the G24 region \citep{2014MNRAS.443.1555U}. In addition, the median mass of the Hi-GAL clump is 850 $M_{\sun}$, which is also higher than that of the CO clumps \citep{2019MNRAS.483.5355M}. The reason for this phenomenon can perhaps be attributed to the higher density of the dust clumps, which usually contain large amounts of gas and dust, and to the fact that a large proportion of the CO clumps in the G24 region are forming small-mass stars rather than massive ones.


\citet{2013ApJ...779..185K} proposed that the smaller the virial parameter $\alpha_{\text{vir}}$, the more dominant the gravitational effect. When the influence of the magnetic field is not taken into account, if $\alpha_{\text{vir}} < 2$ (called the supercritical state, or subvirial state), the clump will collapse and undergo star-forming activity, whereas if $\alpha_{\text{vir}} > 2$ (called the subcritical state, or overvirial state), the clump will tend to expand and thus will not be able to form stars. In the G24 region, we find that most clumps are in gravitationally bound states, as shown in the middle-right panel of Figure \ref{fig:CDF_clumps}, with 48.6\% (125/257), 62.7\% (126/201), and 99.1\% (109/110) of the $^{12}$CO, $^{13}$CO, and C$^{18}$O clumps being in gravitationally bound states, respectively. These ratios are much higher than the proportions of $^{12}$CO, $^{13}$CO, and C$^{18}$O clouds in the MWISP G110 and the G220 regions that are gravitionally bound \citep{2021ApJS..254....3M,2023ApJS..268....1D}. \citet{2014MNRAS.443.1555U} discovered that the vast majority of ATLASGAL MSF clumps have a virial parameter $\alpha_{\text{vir}} < 2$, indicating a gravitationally unstable state. The proportion of gravitationally bound states in MSF clumps is higher than that in CO clumps. Meanwhile, 95.5\% of the Hi-GAL clumps are found to have $\alpha_{\text{vir}} < 2$, with the majority of them having $\alpha_{\text{vir}}$ between 0.1 and 1, suggesting that the Hi-GAL clumps are closer to gravitationally bound states compared to the CO clumps \citep{2019MNRAS.483.5355M}.


The surface densities of the three CO clumps range from 10$^1$ to 10$^3$ $M_{\sun}$ pc$^{-2}$, as seen in the bottom-left panel of Figure \ref{fig:CDF_clumps}, the largest for C$^{18}$O clumps. The median surface densities of $^{12}$CO and $^{13}$CO in the G24 region are an order of magnitude greater than those in the G220 region, whereas the median surface densities of C$^{18}$O are twice as great.


The bottom-right panel of Figure \ref{fig:CDF_clumps} shows that the median number densities of the $^{12}$CO, $^{13}$CO, and C$^{18}$O clumps are 2.79 $\times$ 10$^2$ cm$^{-3}$, 1.31 $\times$ 10$^3$ cm$^{-3}$, 3.91 $\times$ 10$^3$ cm$^{-3}$, respectively, which is consistent with the general understanding that $^{12}$CO is typically used to trace low-density structures, and C$^{18}$O is utilized to trace the densest structures. The median number densities of $^{12}$CO and $^{13}$CO clouds on the different spiral arms in the G110 region are generally smaller than those in the G24 region, reflecting the greater abundance of material in the G24 region \citep{2021ApJS..254....3M}.

\subsubsection{Virial Parameter and Mass of the CO Clumps} \label{subsec:4.1.4}

\begin{figure}[ht]
    \centering
    \includegraphics[width=0.7\linewidth]{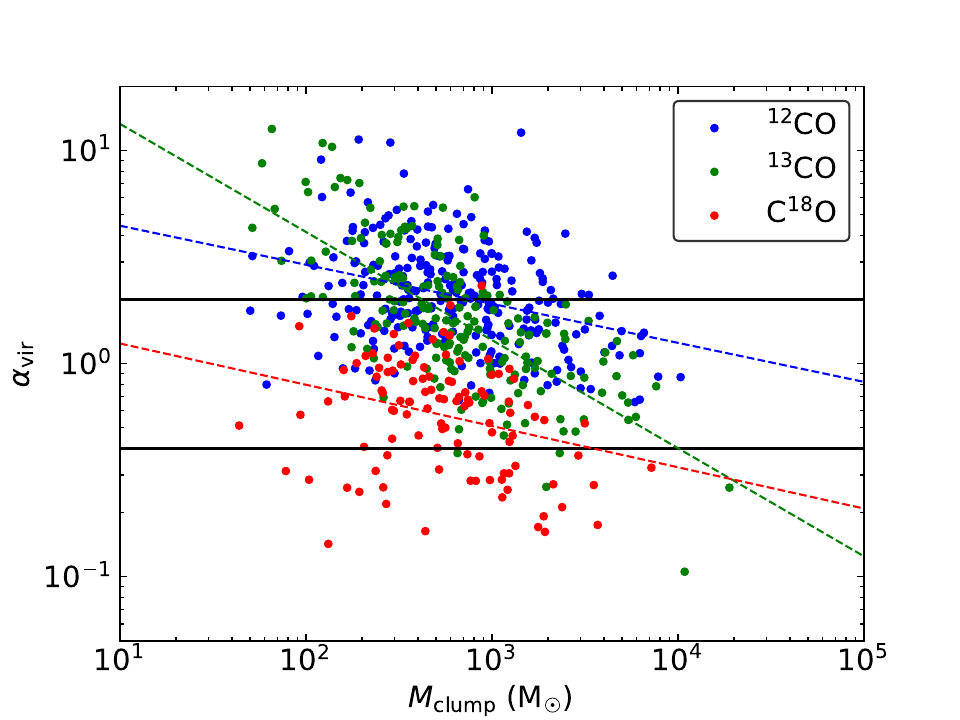}
    \caption{Virial parameter versus clump mass. The two horizontal lines represent $\alpha_{\text{vir}}$ = 2 and $\alpha_{\text{vir}}$ = 0.4. The three dashed lines represent the linear fit to the corresponding clumps.}
    \label{fig:Alpha_vs_Mclump}
\end{figure}

The virial parameter is an important tool for measuring the stability of clumps. The study of the relationship between the virial parameter and the mass of the clumps is significant for understanding the process of star formation. Figure \ref{fig:Alpha_vs_Mclump} illustrates the relationship between the virial parameter and the mass of CO clumps. Following \citet{2013ApJ...779..185K}, we set two horizontal lines at $\alpha_{\text{vir}}$ = 2 and $\alpha_{\text{vir}}$ = 0.4 to demonstrate the stability of the clumps. The horizontal line at $\alpha_{\text{vir}}$ = 2 represents the lowest critical virial parameter for non-magnetised clouds, whereas the horizontal line at $\alpha_{\text{vir}}$ = 0.4 indicates the case where the virial parameter is a factor of 5 lower than $\alpha_{\text{vir}}$ = 2, which further emphasises the instability of clouds without magnetic field support. The latter line provides us with a stricter criterion for identifying clouds that may collapse without the help of a magnetic field. As mentioned earlier, most CO clumps have a virial parameter satisfying $\alpha_{\text{vir}} < 2$ and are therefore in a gravitationally bound state. Studies show that many fragments of molecular cloud clumps have a virial parameter well below 2, with some clumps having a virial parameter as low as 0.4 or even lower \citep{2013ApJ...779..185K}. Within the G24 region, we find that 2.5\% (5/201) of the $^{13}$CO clumps and 28.2\% (31/110) of the C$^{18}$O clumps have a virial parameter below 0.4, suggesting that these clumps are in a supercritical state and may undergo a rapid and violent collapse unless supported by a strong magnetic field. 

Furthermore, the fitting results for $^{12}$CO, $^{13}$CO, and C$^{18}$O clumps are log($\alpha_{\text{vir}}$) = (-0.18 $\pm$ 0.03) $\times$ log($M_{\text{clump}}$) + (0.83 $\pm$ 0.09), log($\alpha_{\text{vir}}$) = (-0.51 $\pm$ 0.04) $\times$ log($M_{\text{clump}}$) + (1.63 $\pm$ 0.10) and log($\alpha_{\text{vir}}$) = (-0.19 $\pm$ 0.06) $\times$ log($M_{\text{clump}}$) + (0.29 $\pm$ 0.17), respectively, with Pearson correlation coefficients $r$ of -0.26, -0.32 and -0.30, which are close to the exponents reported for the G110 and G220 regions \citep{2021ApJS..254....3M,2023ApJS..268....1D}. There is a weak anti-correlation between the virial parameter and clump mass. This trend is also confirmed by whole MSF clumps \citep{2014MNRAS.443.1555U}. As the mass of the clump increases, the virial parameter decreases, suggesting that the more massive the clump, the more likely it is to collapse and form a star.

\subsubsection{Mass and Effective Radius of the CO Clumps} \label{subsec:4.1.5}

\begin{figure}[ht]
    \centering
    \includegraphics[width=0.7\linewidth]{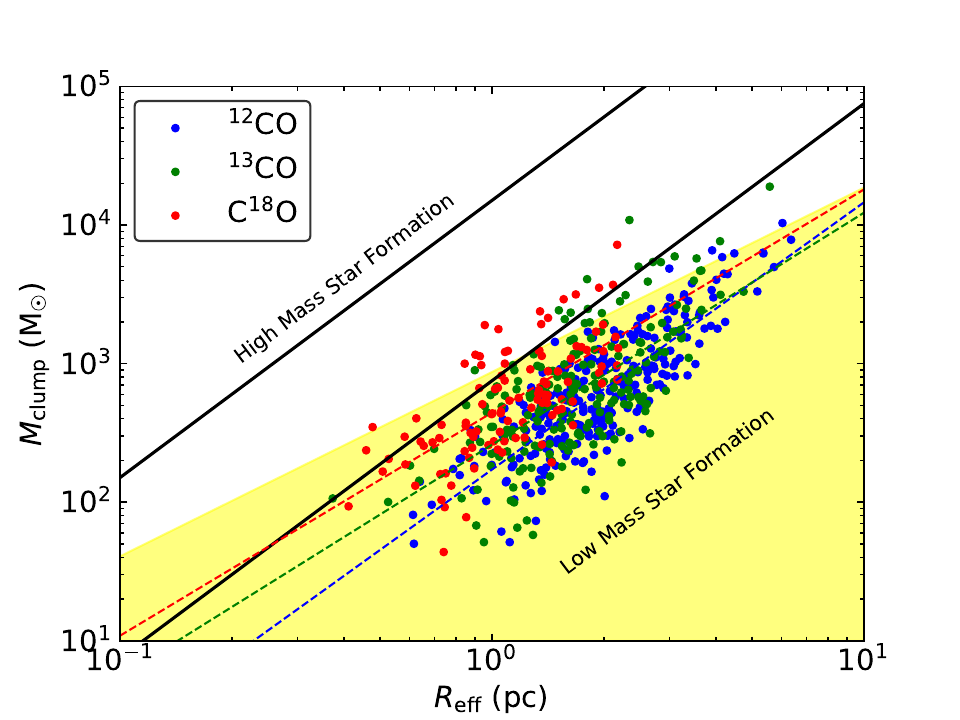}
    \caption{The relationship between the mass and the effective radius of the clump. The upper and lower solid black lines represent surface densities of 1 g cm$^{-2}$ \citep{2008Natur.451.1082K} and 0.05 g cm$^{-2}$ \citep{2013ApJ...779..185K}, respectively. The yellow shaded region shows the region where the mass of the clump is insufficient to form massive stars, and defined as $M(r) \leqslant 870 M_{\sun} (r/\text{pc})^{1.33}$. The three dashed lines represent the linear fit to the corresponding CO clumps.}
    \label{fig:Mclump_vs_Reff}
\end{figure}

In addition to the virial parameter, a common method for determining the potential for clumps to form massive stars is the mass-radius relation. Figure \ref{fig:Mclump_vs_Reff} illustrates the relationship between the clump mass and the effective radius. The upper and lower solid black lines represent surface densities of 1 g cm$^{-2}$ \citep{2008Natur.451.1082K} and 0.05 g cm$^{-2}$ \citep{2013ApJ...779..185K}, respectively, which provide reliable empirical values of the upper and lower bounds on the surface density of clumps required to form massive stars. In addition, \citet{2010ApJ...721..686P} studied the mass-radius relation of neighboring molecular cloud complexes and found that molecular clouds that do not form massive stars typically follow the following formula: $M(r) \leqslant 870 M_{\sun} (r/\text{pc})^{1.33}$, which provides us with another criterion. Combining the above two criteria, we identified 8 and 18 clumps in $^{13}$CO and C$^{18}$O, respectively, with a percentage of 4.0\% (8/201) and 16.4\% (18/110), while none of the clumps in $^{12}$CO meets these two criteria. Figure \ref{fig:Distribution_of_clumps} shows the locations of these clumps with the potential to form massive stars, which are predominantly distributed within the bright CO emission region. The regions of these eight $^{13}$CO clumps, which have the potential to form massive stars, also contain C$^{18}$O clumps with similar potential. A similar situation occurs in the G110 and G220 regions. In the G110 region, there are no $^{12}$CO clouds that meet the conditions for the formation of massive stars, but there are a few $^{13}$CO clouds that do \citep{2021ApJS..254....3M}. For the G220 region, neither $^{12}$CO, $^{13}$CO nor C$^{18}$O clouds fulfill the conditions for the formation of massive stars \citep{2023ApJS..268....1D}. The same analysis was also performed by \citet{2014MNRAS.443.1555U}, where almost all MSF clumps satisfy the conditions for the formation of massive stars, suggesting that MSF clumps are more likely to form massive stars than CO clumps. A similar result holds for Hi-GAL sources, most of which have the capability to form massive stars \citep{2017MNRAS.471..100E,2021MNRAS.504.2742E}.


We performed a linear fitting to the log-log relations of the mass and effective radius of the clumps. The linear fitting results are log($M_{\text{clump}}$) = (1.93 $\pm$ 0.08) $\times$ log($R_{\text{eff}}$) + (2.24 $\pm$ 0.03), log($M_{\text{clump}}$) = (1.67 $\pm$ 0.13) $\times$ log($R_{\text{eff}}$) + (2.41 $\pm$ 0.04) and log($M_{\text{clump}}$) = (1.61 $\pm$ 0.16) $\times$ log($R_{\text{eff}}$) + (2.65 $\pm$ 0.03) for $^{12}$CO, $^{13}$CO and C$^{18}$O clumps, respectively, with Pearson correlation coefficients, $r$, of 0.83, 0.63 and 0.56, respectively, indicating a significant positive correlation between the mass and radius of the clumps. Both \citet{2021ApJS..254....3M} and \citet{2023ApJS..268....1D} obtained an exponent of $\sim$ 2.2 for the power-law function of the $^{12}$CO cloud and $\sim$ 2.4 for the power-law exponent of the $^{13}$CO cloud in relation $M-R$, which is steeper than the fitted curves in the G24 region. \citet{2014MNRAS.443.1555U} similarly found a strong positive correlation between these two parameters in ATLASGAL MSF clumps, with a correlation coefficient of 0.85 and a logarithmic relationship ($M_{\text{clump}}$) = (1.67 $\pm$ 0.03) $\times$ log($R_{\text{eff}}$) + (3.42 $\pm$ 0.01), which is in very good agreement with our results.

\subsubsection{Velocity Dispersion - Size Relation} \label{subsec:4.1.6}

The velocity dispersion ($\sigma_v$) is an important parameter describing the motion of the gas inside the molecular cloud, reflecting the level of disordered motion of the gas, i.e., the intensity of turbulent motion. Turbulence plays a dominant role in large-scale molecular clouds, both by suppressing gravitational collapse within the cloud and by driving the formation of complex density structures such as filamentary structures and denser clumps/cores, which are sites of star formation \citep{2020ApJ...900...82P}. \citet{1981MNRAS.194..809L} found a power-law relationship between the velocity dispersion $\sigma_v$ of the gas inside a molecular cloud and the radius of the cloud $R$, i.e., $\sigma_v \varpropto R^{0.38}$, which is subsequently modified to $\sigma_v \varpropto R^{0.5}$\citep{1987ApJ...319..730S}. This relationship is known as Larson's velocity dispersion - size relation.

\begin{figure}[ht]
    \centering
    \includegraphics[width=0.7\linewidth]{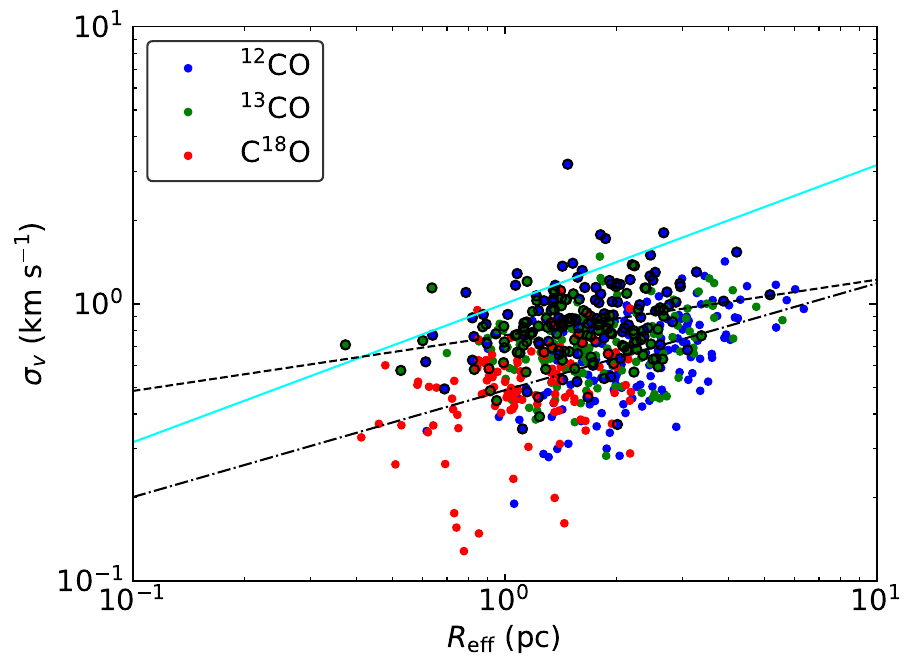}
    \caption{Velocity dispersion as a function of effective radius. CO and its isotopic clumps with $\alpha_{\text{vir}} > 2$ are points labelled with black circles and linearly fitted with black dashed line, while clumps with $\alpha_{\text{vir}} \leqslant 2$ are unlabelled points and linearly fitted with black dot-dashed line. The cyan solid line is the referenced line of the Larson relation $\sigma \varpropto R^{0.5}$.}
    \label{fig:sigmav_vs_Reff}
\end{figure}

In this study, we classify the clumps into two categories based on the threshold $\alpha_{\text{vir}} = 2$: the overvirial state ($\alpha_{\text{vir}} > 2$) and the subvirial state ($\alpha_{\text{vir}} \leqslant 2$), as shown in Figure \ref{fig:sigmav_vs_Reff}. The Pearson correlation coefficients $r$ for the overvirial clumps and the subvirial clumps are 0.23 and 0.54, respectively. By linear fitting, we obtain the following results: the fitting equation is log($\sigma_v$) = (0.20 $\pm$ 0.05) $\times$ log($R_{\text{eff}}$) + (-0.11 $\pm$ 0.01) for the overvirial clumps, and log($\sigma_v$) = (0.39 $\pm$ 0.03) $\times$ log($R_{\text{eff}}$) + (-0.31 $\pm$ 0.01) for the subvirial clumps. The results show that the slope of the overvirial clumps (0.20) is much lower than the standard value of the Larson relation (0.5), while the slope of the subvirial clumps (0.37) is much closer to the initial power exponent of the Larson relation (0.38). Throughout the scale range (0.1 to 10 pc), the velocity dispersion of the overvirial clumps is larger than that of the subvirial clumps. At the same time, the velocity dispersion of the subvirial clumps is about half that of the Larson relation, suggesting that the turbulent motion inside the subvirial clumps is significantly weaker than that of the overall molecular cloud at the same effective radius, and thus gravity dominates in the subvirial clumps, thus contributing to their collapse process.

\subsubsection{\texorpdfstring{$\sigma_{v}/R^{0.5}$ - $\Sigma$ Relation}{sigma\_v/R0.5-Sigma Relation}} \label{subsec:4.1.7}

\begin{figure}[ht]
    \centering
    \includegraphics[width=0.7\linewidth]{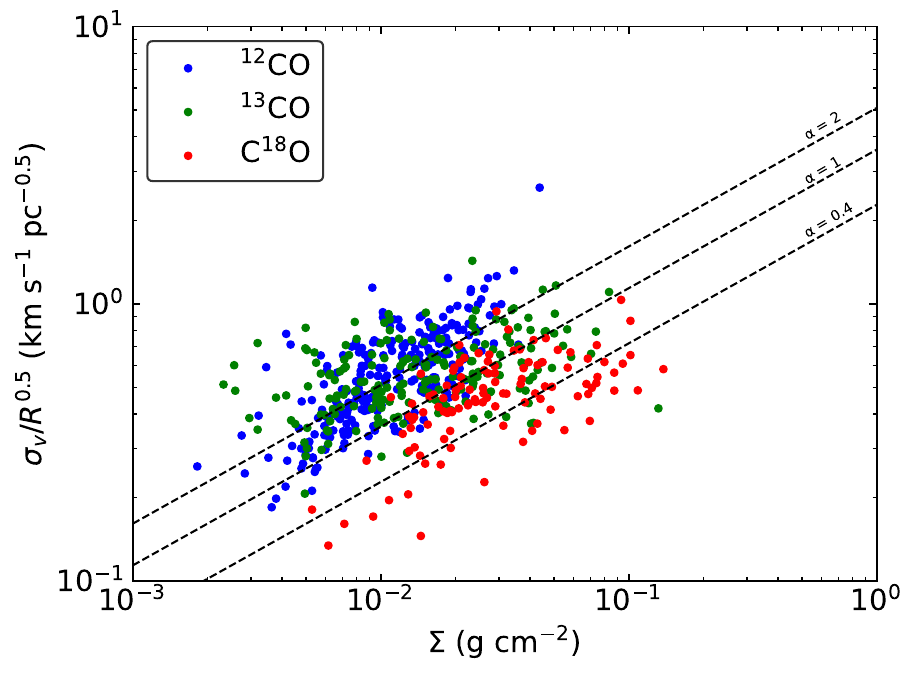}
    \caption{Variation of $\sigma_{v}/R^{0.5}$ with surface density $\Sigma$. The three black dashed lines mark the positions at which the virial coefficients $\alpha_{\text{vir}}$ are 2, 1, and 0.4, respectively.}
    \label{fig:sigmavR0.5_vs_Sigma}
\end{figure}

The dynamics of the molecular cloud can be characterized by the relation $\sigma_{v}/R^{0.5} \varpropto \Sigma^{0.5}$ \citep{2015ARA&A..53..583H}. This relationship is a more generalized form of the Larson scaling relationship for cases where the surface density $\Sigma$, is not constant \citep{2016ApJ...833..113C}. This relation indicates that the cloud is in a state where the gravitational potential energy and the kinetic energy are approximately equal, in accordance with Viry's theorem, which suggests that these molecular clouds are gravity-bound stable structures \citep{2009ApJ...699.1092H}. \citet{2011MNRAS.411...65B} proposed that this relation is valid due to a general gravitational collapse, rather than the cloud being in dynamical equilibrium. As shown in Figure \ref{fig:sigmavR0.5_vs_Sigma}, $\sigma_{v}/R^{0.5}$ shows a relatively clear positive correlation with the cloud surface density $\Sigma$, with Pearson correlation coefficients $r$ of 0.71, 0.37 and 0.49 for $^{12}$CO, $^{13}$CO and C$^{18}$O clumps, respectively, which suggests that the molecular cloud is in virial equilibrium. For the Hi-GAL source, its $\sigma_{v}/R^{0.5}$ parameter also shows a weak positive correlation with the surface density $\Sigma$, in agreement with the trend of our results \citep{2019MNRAS.483.5355M}. As clouds become smaller and denser, this trend corresponds to a transition from turbulence-dominated cloud structures ($\alpha_{\text{vir}} > 2$) to gravitationally-dominated cloud structures ($\alpha_{\text{vir}} \leqslant 2$), suggesting that gravity begins to dominate the dynamical processes of the clouds on smaller scales \citep{2024AJ....167..228L}.

\subsection{Gas Kinematics} \label{sec:4.2}

Infall motion is the process by which materials such as gas and dust gradually gather and collapse under the influence of gravity, which plays a key role in star formation \citep{1987ARA&A..25...23S}. The method of identifying infall clumps is by verifying spectral line profiles \citep{1993ApJ...404..232Z,1997ApJ...489..719M,2023RAA....23g5001J}. Models of infall clumps (e.g., \citealt{1977ApJ...214L..73L}) suggest that infalling gas during star formation leads to red-shifted self-absorbing profiles of optically thick lines, forming characteristic blue asymmetric profiles, including double-peak profiles, peak-shoulder profiles, and single-peak profiles with the peak skewed toward the blue-shifted side, which are uniformly categorized as blue profiles \citep{2020RAA....20..115Y,2021ApJ...922..144Y}. In addition to infall motion, another common reason that contributes to the formation of blue profiles is the presence of a multi-velocity component in the line of sight. In order to effectively distinguish between these two cases, we use single-peak optical thin lines without self-absorption to track the central velocity of the clump. If the peak velocity of the optically thin line lies between the blue and red peaks of the optically thick line, we can assume that the clump is undergoing infall \citep{2023ApJ...955..154Y,2024AJ....168...52Y}. In this study, the HCO$^{+}$ line was chosen for the optically thick line because it is more effective at tracing the infall motion compared to the $^{12}$CO line, mainly due to the higher critical density of HCO$^{+}$ \citep{2022RAA....22i5014Y}. Meanwhile, the C$^{18}$O line was chosen for the optically thin line.

\begin{figure}[ht]
    \centering
    \includegraphics[width=0.45\linewidth]{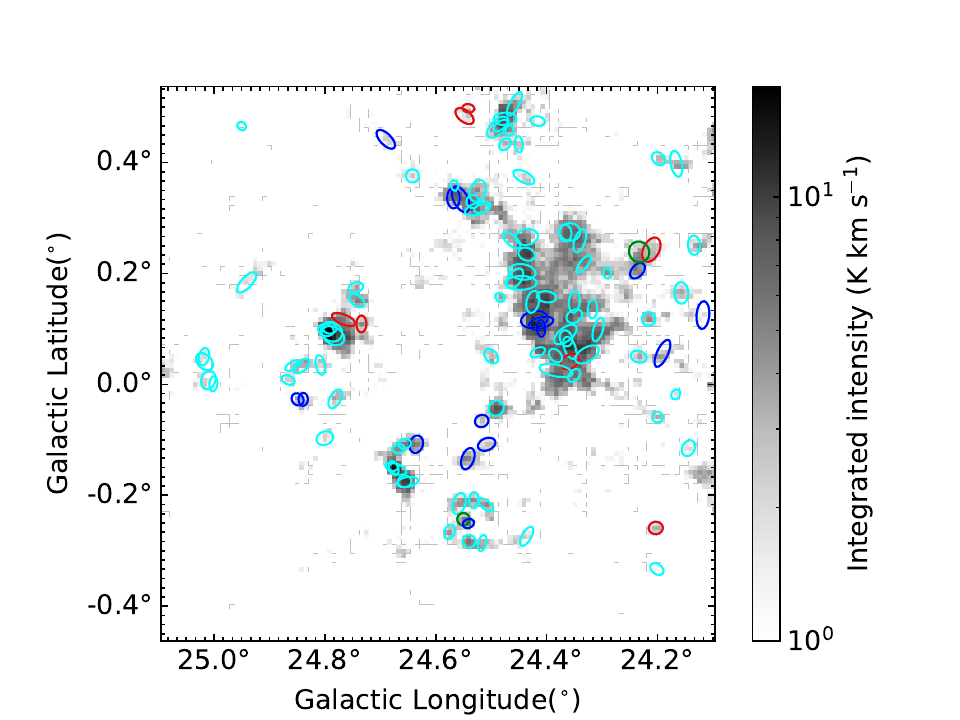}
    \caption{Spatial distribution of the infall candidates identified from the HCO$^{+}$ line profiles. The background is the integrated intensity map of C$^{18}$O emission. The ellipses indicate the location of the C$^{18}$O clumps, where cyan colored ones indicate the locations of the C$^{18}$O clumps with non-blue HCO$^{+}$ profiles, while red, green and blue colored ones indicate the locations of the C$^{18}$O clumps with double-peak HCO$^{+}$, peak-shoulder HCO$^{+}$, and single-peak HCO$^{+}$ profiles with the peak skewed toward the blue, respectively. }
    \label{fig:Distribution_of_infall}
\end{figure}
\begin{figure}[htp]
    \centering
    \includegraphics[width=0.75\linewidth]{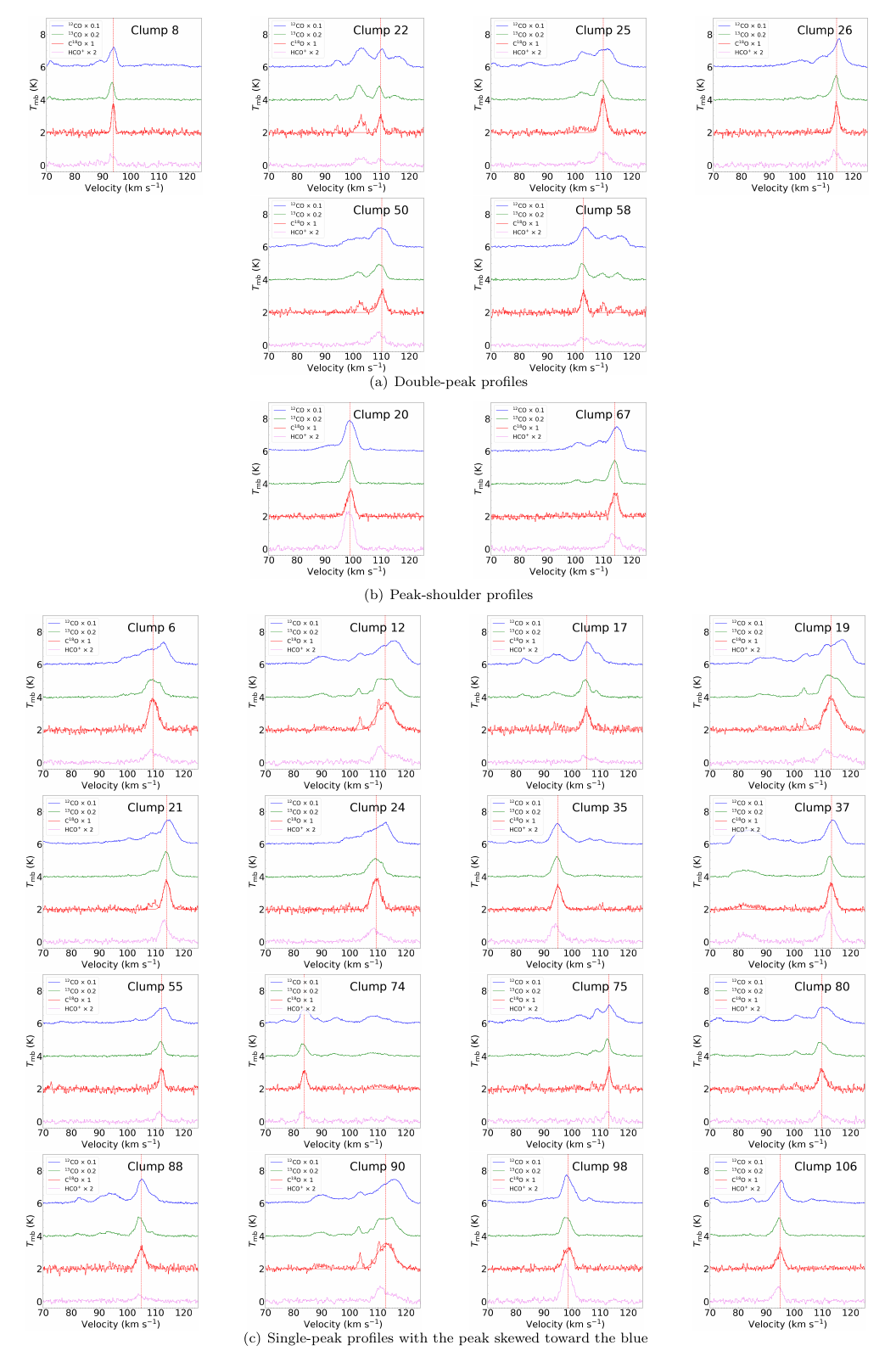}
    \caption{The $^{12}$CO, $^{13}$CO, C$^{18}$O and HCO$^{+}$ line profiles towards the infall candidates. The red dashed line indicates the center velocity of the C$^{18}$O line obtained by Gaussian fitting. Clump numbers indicate the serial numbers of the C$^{18}$O clumps, see Table \ref{appendix:the_parameters_of_C18O_clumps}.}
    \label{fig:Infall_spectrum}
\end{figure}

We utilized C$^{18}$O clumps for the study of infall motion, the spatial distribution of which is shown in Figure \ref{fig:Distribution_of_infall}. 
Since the G24 region is an active star-forming region, the velocity components of the CO and HCO$^{+}$ lines of parts of the clumps are very complicated, making it impossible to judge whether or not the blue profiles originate from infall motions. Therefore, we only selected C$^{18}$O clumps with relatively isolated emission components of CO and HCO$^{+}$ lines for an in-depth study. Of the 110 C$^{18}$O clumps, the final number of valid clumps used to find the infall spectral line was 106 because four of them missed HCO$^{+}$ line data. Among the 106 C$^{18}$O clumps, we identified 6 double-peak profiles, 2 peak-shoulder profiles, and 16 single-peak profiles with the peak skewed toward the blue. The specific line profiles of these infall candidates are presented in Figure \ref{fig:Infall_spectrum}.

The infall velocity of the infall clumps can help us better understand the dynamical behavior of molecular clouds. A common method for estimating the infall velocity is that it can be crudely estimated by:

\begin{equation}
    V_{\text{in}} = V_{\text{sys}} - V_{\text{blue}}, 
\end{equation}
where $V_{\text{sys}}$ is the systematic velocity obtained by Gaussian fitting of the optically thin line C$^{18}$O, and $V_{\text{blue}}$ represents the blue-peak velocity of the optically thick line HCO$^{+}$.


The estimated infall velocity $V_{\text{in}}$ is shown in Table \ref{tab:Vin and Min}. Infall velocities $V_{\text{in}}$ in a range of 0.37 to 2.18 km s$^{-1}$, with a median value of 0.91 km s$^{-1}$ in our observed regions. This result is consistent with previous studies \citep{2013A&A...549A...5R,2016MNRAS.456.2681Q} on MSFRs. \citet{2022RAA....22i5014Y} summarized the literature on infall motions and found that $V_{\text{in}}$ is usually less than 0.5 km s$^{-1}$ for low-mass clumps, while that of high-mass clumps is more than 1 km s$^{-1}$. According to this criterion, our results show that 4.2\% (1/24) of the infall clumps are classified as low-mass, while 37.5\% (9/24) of the infall clumps belong to high-mass. This finding suggests that one-third of the infall clumps have the potential to form massive stars in our observed region.

\begin{table}[ht]
\centering
\caption{Infall velocity, mass infall rate of infall clumps}
\begin{tabular}{ccc}
\hline
\hline
Source         & $V_{\text{in}}$  & $\dot{M}_{\text{in}}$                     \\
               & (km s$^{-1}$)    & ($\times 10^{-4} M_{\sun}$ yr$^{-1}$)     \\
(1)            & (2)              & (3)                                       \\
\hline
\multicolumn{3}{c}{Double-peak profiles}                                      \\
\hline
8              & 1.01             & 6.0                                       \\
22             & 0.61             & 3.9                                       \\
25             & 1.35             & 58.8                                      \\
26             & 1.04             & 14.3                                      \\
50             & 0.82             & 10.8                                      \\
58             & 0.98             & 7.2                                       \\
\hline
\multicolumn{3}{c}{Peak-shoulder profiles}                                    \\
\hline
20             & 0.52             & 6.7                                       \\
67             & 1.06             & 11.4                                      \\
\hline
\multicolumn{3}{c}{Single-peak profiles with the peak skewed toward the blue} \\
\hline
6              & 0.89             & 24.4                                      \\
12             & 1.64             & 75.9                                      \\
17             & 1.12             & 8.3                                       \\
19             & 2.18             & 221.6                                     \\
21             & 0.67             & 11.1                                      \\
24             & 0.64             & 7.8                                       \\
35             & 0.87             & 19.0                                      \\
37             & 0.79             & 15.1                                      \\
55             & 0.74             & 40.3                                      \\
74             & 0.99             & 8.3                                       \\
75             & 0.68             & 9.2                                       \\
80             & 0.89             & 19.1                                      \\
88             & 1.11             & 7.5                                       \\
90             & 2.05             & 107.1                                     \\
98             & 0.92             & 14.4                                      \\
106            & 0.37             & 4.0                                       \\
\hline
\end{tabular}
\tablecomments{Column (1): serial number of the C$^{18}$O clump. Column (2): falling velocity. Column (3): mass falling rate.}
\label{tab:Vin and Min}
\end{table}

The mass-infall rate $\dot{M}_{\text{in}}$ can be roughly obtained by the following formula \citep{2010A&A...517A..66L}:

\begin{equation}
    \dot{M}_{\text{in}} = 4 \pi R^2 V_{\text{in}} \rho,
\end{equation}
where $R$ is the effective radius of the clump, $V_{\text{in}}$ is the infall velocity, and $\rho$ is the volume density of the C$^{18}$O clump. 
Mass-infall rates are detailed in Table \ref{tab:Vin and Min}, with a range of (3.9 -- 221.6) $\times 10^{-4} M_{\sun}$ yr$^{-1}$ and a median of 11.3 $\times 10^{-4} M_{\sun}$ yr$^{-1}$. According to prior research, low-mass star formation is characterized by mass-infall rates between 10$^{-6}$ and 10$^{-5}$ $M_{\sun}$ yr$^{-1}$ \citep{2013A&A...549A...5R,2021ApJ...910..112K}, whereas massive star formation is associated with clumps exhibiting mass-infall rates between 10$^{-4}$ and 10$^{-2}$ $M_{\sun}$ yr$^{-1}$ \citep{2005MNRAS.360.1506K,2020ApJ...904..181L}. Our findings indicate that the mass-infall rates of all identified infall clumps surpass 10$^{-4}$ $M_{\sun}$ yr$^{-1}$, implying that they are likely to be forming massive stars. 


\citet{2024AJ....168...52Y} proposed that sources with peak-shoulder profiles exhibit slightly higher infall velocities and mass-infall rates compared to those with double-peak profiles. In our study, the average infall velocities for clumps presenting double-peak profiles, peak-shoulder profiles, and single-peak profiles skewed towards the blue are 0.97 km s$^{-1}$, 0.80 km s$^{-1}$, and 1.04 km s$^{-1}$, respectively. Consequently, the average mass-infall rates are 16.8 $\times 10^{-4} M_{\sun}$ yr$^{-1}$, 9.1 $\times 10^{-4} M_{\sun}$ yr$^{-1}$, and 37.1 $\times 10^{-4} M_{\sun}$ yr$^{-1}$, respectively. The infall velocities and mass-infall rates of the peak-shoulder sources are both lower than those of the double-peak sources, which deviates from the findings of \citet{2024AJ....168...52Y}. This discrepancy may be attributed to the smaller number of peak-shoulder sources. However, both the infall velocities and mass-infall rates of the single-peak sources with the peak skewed towards the blue are higher than those of the double-peak sources, indicating that the single-peak sources with a blue-skewed peak tend to form stars of greater mass.

In addition to examining infall motions, we have also sought outflow phenomena by analyzing the wings of the $^{12}$CO molecular line relative to C$^{18}$O clumps. However, no significant CO outflow structures were identified around the C$^{18}$O clumps. This lack of detection can be attributed to several factors. The G24 region's proximity to the Milky Way's galactic center results in a complex interstellar environment that obscures the clear identification of potential outflow clumps. The influence of strong gravitational and magnetic fields, coupled with intense background radiation, likely inhibits the formation of outflows. Moreover, outflows are known to occur at specific evolutionary stages of star formation; if a significant number of clumps in the G24 region are still in the early stages of star formation or have not yet reached a stage where they can drive outflows, this could account for the observed paucity of outflow events. Most importantly, the intricate gas dynamics and large-scale gas motions associated with the rotation of the Galactic bar in the G24 region may be the predominant factors contributing to the minimal detection of outflows.

\section{Summary} \label{sec:5}

The G24 region is an active star-forming region in the Milky Way, located at intersection of the Norma arm, the 3 kpc arm, and the near-end of the Galactic bar, exhibiting a complex molecular cloud structure and multiple potentially massive star-forming clumps. By observing the molecular clouds in this region, it would be able to better understand their physical properties and star formation processes. Therefore, in this paper, the distributions and physical properties of molecular clouds in the G24 region (1 square degree) are investigated with molecular lines detected by the PMO 13.7 m telescope. In conjunction with the MWISP project and the OTF mapping, we obtained observational data of $^{12}$CO, $^{13}$CO, C$^{18}$O, and HCO$^{+}$ (1-0). The main findings and conclusions are summarized below.


(1) The physical properties of the CO molecular cloud in seven subregions of the G24 region were investigated, including excitation temperature, column density, mass and surface density. The LTE method and the X-factor method were used to estimate the H$_2$ column density, and the $X_{\text{CO}}$ of the G24 region is 8.25 $\times$ 10$^{19}$ cm$^{-2}$ (K km s$^{-1}$)$^{-1}$. Among them, subregion E has significantly larger $T_{\text{pk}}$, $T_{\text{ex}}$, $N_{\text{H$_2$}}$, $M$, and $\Sigma$ than the other subregions, suggesting that the subregion is very active in star formation. 


(2) The GaussClumps algorithm was used to identify the clumps in the molecular cloud, and a total of 257, 201 and 110 were identified from the $^{12}$CO, $^{13}$CO and C$^{18}$O molecules, respectively. The physical parameters of these clumps were calculated, including excitation temperature, column density, effective radius, clump mass, and virial parameter. Based on the criterion of the virial parameter $\alpha_{\text{vir}} \leq 2$, the majority of the clumps are gravitationally bound, with gravitationally bound ratios of 48.6\%, 62.7\%, and 99.1\% for $^{12}$CO, $^{13}$CO and C$^{18}$O clumps, respectively. Among the $^{12}$CO, $^{13}$CO and C$^{18}$O clumps, the number with the potential to form massive stars is 0, 8, and 18, respectively, with proportions of $^{13}$CO and C$^{18}$O accounting for 4.0\%, and 16.4\% of the total number of clumps.


(3) Dynamical processes within the G24 region, such as the infall and outflow of clumps, have been thoroughly examined. Through the analysis of HCO$^{+}$ and C$^{18}$O spectral line profiles, we identified six double-peak profiles, two peak-shoulder profiles, and sixteen single-peak profiles with a blue-skewed peak. The substantial infall velocities (with a median of 0.91 km s$^{-1}$) and mass infall rates (with a median of 11.3 $\times 10^{-4} M_{\sun}$ yr$^{-1}$) of these infall clumps indicate a high likelihood of massive star formation in the G24 region. In addition, single-peak profiles with a blue-skewed peak exhibit significantly higher infall velocities and mass-infall rates compared to double-peak profiles, suggesting a greater propensity for massive star formation within this type of infall clumps. However, gas outflow activities in the G24 region are seldom observed, possibly due to complex gas kinematics arising from intricate molecular cloud structures or the overarching gas motions associated with the Galactic bar's rotation, which may obscure the detection of gas outflows.

\acknowledgments
We thank the PMO for providing the MWISP data and the operator's assistance during the observations, as well as Yi-Wei Dong from the PMO for her help and advice. This work is supported by the National Key R\&D Program of China (2022YFA1603102) and the National Natural Science Foundation of China (12473024) and (12403026). This work is also supported by the Guangzhou University Postgraduate Innovation Ability Cultivation Programme. X.C. thanks Guangdong Province Universities and Colleges Pearl River Scholar Funded Scheme (2019).

\clearpage
\appendix
\section{The parameters of clumps \label{appendix:the physical properties of clumps}}
This appendix lists the measured physical parameters of 257 $^{12}$CO, 201 $^{13}$CO, and 110 C$^{18}$O clumps.

\setcounter{table}{0} 
\renewcommand{\thetable}{A\arabic{table}} 

\begin{longrotatetable}
\addtolength{\tabcolsep}{-1.5pt}
\begin{deluxetable*}{ccccccccccccccccccccc}
\tablecaption{The parameters of $^{12}$CO clumps}
\tabletypesize{\scriptsize}
\tablehead{
\colhead{Number} & \colhead{Source} & \colhead{$V_{\text{lsr}}$} & \colhead{$\Delta v$} & \colhead{$\theta_{\text{a}}$} & \colhead{$\theta_{\text{b}}$} & \colhead{$\theta_{\text{fwhm}}$} & \colhead{P.A.} & \colhead{Distance} & \colhead{$R_{\text{GC}}$} & \colhead{$T_{\text{ex}}$} & \colhead{log($N_{\text{H$_2$}}$)} & \colhead{$R_{\text{eff}}$} & \colhead{log($M_{\text{clump}}$)} & \colhead{log($M_{\text{vir}}$)} & \colhead{$\alpha_{\text{vir}}$} & \colhead{log($\Sigma$)} & \colhead{log($n_{\text{H$_2$}}$)} & \colhead{$\Delta v_{\text{th}}$} & \colhead{$\Delta v_{\text{nth}}$} & \colhead{Potential} \\
\colhead{} & \colhead{} & \colhead{(km s$^{-1}$)} & \colhead{(km s$^{-1}$)} & \colhead{($\arcsec$)} & \colhead{($\arcsec$)} & \colhead{($\arcsec$)} & \colhead{($\arcdeg$)} & \colhead{(kpc)} & \colhead{(kpc)} & \colhead{(K)} & \colhead{(cm$^{-2}$)} & \colhead{(pc)} & \colhead{($M_{\sun}$)} & \colhead{($M_{\sun}$)} & \colhead{} & \colhead{($M_{\sun}$ pc$^{-2}$)} & \colhead{(cm$^{-3}$)} & \colhead{(km s$^{-1}$)} & \colhead{(km s$^{-1}$)} & \colhead{}
\decimalcolnumbers
}
\startdata
1  & MWISP G24.469$+$0.483 & 103.48 & 4.05 & 126 & 167 & 142 & 75  & 5.66 & 3.81 & 14.3 & 21.6 & 1.9 & 3.2 & 3.8 & 4.2 & 2.1 & 2.4 & 0.15 & 4.05 &  \\
2  & MWISP G24.506$-$0.234 & 99.83  & 2.94 & 91  & 167 & 113 & 115 & 5.71 & 3.79 & 15.4 & 21.5 & 1.6 & 3.0 & 3.5 & 3.1 & 2.1 & 2.2 & 0.16 & 2.94 &  \\
3  & MWISP G24.458$+$0.502 & 100.63 & 2.39 & 142 & 172 & 155 & 10  & 5.59 & 3.84 & 12.8 & 21.3 & 2.0 & 3.0 & 3.4 & 2.7 & 1.8 & 1.8 & 0.14 & 2.39 &  \\
4  & MWISP G24.555$-$0.235 & 98.56  & 2.05 & 128 & 202 & 152 & 116 & 5.67 & 3.81 & 15.1 & 21.3 & 2.1 & 3.0 & 3.3 & 1.8 & 1.9 & 1.7 & 0.16 & 2.04 &  \\
5  & MWISP G24.670$-$0.156 & 113.17 & 2.97 & 97  & 244 & 128 & 108 & 6.94 & 3.43 & 14.0 & 21.4 & 2.5 & 3.3 & 3.7 & 2.4 & 2.0 & 1.9 & 0.15 & 2.97 &  \\
6  & MWISP G24.788$+$0.094 & 112.34 & 2.19 & 106 & 150 & 122 & 118 & 6.63 & 3.50 & 17.0 & 21.4 & 1.9 & 3.0 & 3.3 & 1.7 & 2.0 & 1.8 & 0.17 & 2.18 &  \\
7  & MWISP G24.723$+$0.081 & 109.85 & 1.94 & 66  & 115 & 81  & -11 & 6.42 & 3.55 & 13.4 & 21.2 & 1.1 & 2.5 & 2.9 & 2.8 & 1.9 & 2.2 & 0.15 & 1.93 &  \\
8  & MWISP G24.468$+$0.187 & 119.20 & 2.19 & 89  & 129 & 103 & 57  & 7.23 & 3.38 & 11.1 & 21.2 & 1.7 & 2.7 & 3.2 & 3.2 & 1.8 & 1.9 & 0.13 & 2.19 &  \\
9  & MWISP G24.537$-$0.210 & 101.26 & 2.75 & 72  & 167 & 93  & 10  & 5.78 & 3.76 & 18.5 & 21.6 & 1.4 & 2.9 & 3.3 & 2.5 & 2.2 & 2.3 & 0.17 & 2.74 &  \\
10 & MWISP G24.764$+$0.072 & 110.28 & 2.08 & 161 & 239 & 189 & 156 & 6.44 & 3.55 & 18.3 & 21.4 & 3.0 & 3.4 & 3.4 & 1.0 & 2.0 & 1.4 & 0.17 & 2.07 &  \\
\enddata
\tablecomments{Column (1): clump number. Column (2): source name defined by the Galactic Coordinates. Columns (3) and (4): central velocity and linewidth of each clump. Columns (5) -- (8): major axis, minor axis, diameter and positional angle of the clumps. Columns (9) and (10): heliocentric distance and Galactocentric distance. Column (11): excitation temperature. Column (12): hydrogen column density. Column (13): effective radius. Column (14): mass of clumps. Column (15): virial mass. Column (16): virial parameter. Column (17): surface density. Column (18): number density of hydrogen molecules. Columns (19) and (20): thermal and non-thermal linewidth. Columns (21): whether the clumps have the potential to form massive stars (details see subsection \ref{subsec:4.1.5}).\\
(This table is available in its entirety in machine-readable form.)}
\label{appendix:the_parameters_of_12CO_clumps}
\end{deluxetable*}
\end{longrotatetable}

\begin{longrotatetable}
\addtolength{\tabcolsep}{-1.5pt}
\begin{deluxetable*}{ccccccccccccccccccccc}
\tablecaption{The parameters of $^{13}$CO clumps}
\tabletypesize{\scriptsize}
\tablehead{
\colhead{Number} & \colhead{Source} & \colhead{$V_{\text{lsr}}$} & \colhead{$\Delta v$} & \colhead{$\theta_{\text{a}}$} & \colhead{$\theta_{\text{b}}$} & \colhead{$\theta_{\text{fwhm}}$} & \colhead{P.A.} & \colhead{Distance} & \colhead{$R_{\text{GC}}$} & \colhead{$T_{\text{ex}}$} & \colhead{log($N_{\text{H$_2$}}$)} & \colhead{$R_{\text{eff}}$} & \colhead{log($M_{\text{clump}}$)} & \colhead{log($M_{\text{vir}}$)} & \colhead{$\alpha_{\text{vir}}$} & \colhead{log($\Sigma$)} & \colhead{log($n_{\text{H$_2$}}$)} & \colhead{$\Delta v_{\text{th}}$} & \colhead{$\Delta v_{\text{nth}}$} & \colhead{Potential} \\
\colhead{} & \colhead{} & \colhead{(km s$^{-1}$)} & \colhead{(km s$^{-1}$)} & \colhead{($\arcsec$)} & \colhead{($\arcsec$)} & \colhead{($\arcsec$)} & \colhead{($\arcdeg$)} & \colhead{(kpc)} & \colhead{(kpc)} & \colhead{(K)} & \colhead{(cm$^{-2}$)} & \colhead{(pc)} & \colhead{($M_{\sun}$)} & \colhead{($M_{\sun}$)} & \colhead{} & \colhead{($M_{\sun}$ pc$^{-2}$)} & \colhead{(cm$^{-3}$)} & \colhead{(km s$^{-1}$)} & \colhead{(km s$^{-1}$)} & \colhead{}
\decimalcolnumbers
}
\startdata
1  & MWISP G24.473$+$0.480 & 102.38 & 2.84 & 108 & 191 & 133 & 83  & 5.64 & 3.82 & 13.8 & 21.7 & 1.8 & 3.3 & 3.5 & 1.6 & 2.3 & 3.6 & 0.15 & 2.84 &  \\
2  & MWISP G24.670$-$0.156 & 112.34 & 2.54 & 84  & 213 & 111 & 112 & 6.69 & 3.48 & 14.0 & 21.6 & 2.0 & 3.3 & 3.4 & 1.4 & 2.2 & 3.4 & 0.15 & 2.54 &  \\
3  & MWISP G24.780$+$0.088 & 110.13 & 2.28 & 154 & 215 & 177 & 129 & 6.40 & 3.56 & 18.7 & 21.8 & 2.7 & 3.7 & 3.5 & 0.5 & 2.4 & 3.5 & 0.17 & 2.27 &  \\
4  & MWISP G24.454$+$0.191 & 118.82 & 2.02 & 100 & 145 & 116 & 39  & 7.23 & 3.38 & 10.6 & 21.2 & 1.9 & 2.8 & 3.2 & 2.4 & 1.8 & 3.0 & 0.13 & 2.02 &  \\
5  & MWISP G24.490$-$0.042 & 110.49 & 2.56 & 99  & 116 & 107 & 86  & 6.68 & 3.46 & 10.2 & 21.1 & 1.5 & 2.6 & 3.3 & 5.5 & 1.7 & 3.1 & 0.13 & 2.56 &  \\
6  & MWISP G24.559$-$0.217 & 99.67  & 1.50 & 131 & 131 & 131 & 63  & 5.73 & 3.78 & 15.5 & 21.3 & 1.7 & 2.9 & 2.9 & 1.1 & 1.9 & 3.2 & 0.16 & 1.49 &  \\
7  & MWISP G24.552$-$0.246 & 98.22  & 1.88 & 80  & 117 & 94  & 39  & 5.65 & 3.82 & 12.9 & 21.3 & 1.1 & 2.5 & 2.9 & 2.5 & 1.9 & 3.4 & 0.14 & 1.87 &  \\
8  & MWISP G24.232$+$0.227 & 113.83 & 2.10 & 159 & 292 & 197 & 46  & 7.21 & 3.35 & 10.2 & 21.2 & 3.7 & 3.4 & 3.5 & 1.4 & 1.8 & 2.7 & 0.13 & 2.10 &  \\
9  & MWISP G24.445$+$0.506 & 100.88 & 1.83 & 91  & 98  & 94  & -26 & 5.60 & 3.83 & 13.9 & 21.3 & 1.1 & 2.5 & 2.9 & 2.5 & 1.9 & 3.4 & 0.15 & 1.82 &  \\
10 & MWISP G24.436$+$0.260 & 117.01 & 1.75 & 106 & 193 & 132 & 97  & 7.16 & 3.38 & 9.9  & 21.0 & 2.3 & 2.8 & 3.2 & 2.3 & 1.6 & 2.8 & 0.13 & 1.75 &  \\
\enddata
\tablecomments{Columns are same as Table \ref{appendix:the_parameters_of_12CO_clumps}.\\
(This table is available in its entirety in machine-readable form.)}
\label{appendix:the_parameters_of_13CO_clumps}
\end{deluxetable*}
\end{longrotatetable}

\begin{longrotatetable}
\addtolength{\tabcolsep}{-1.5pt}
\begin{deluxetable*}{ccccccccccccccccccccc}
\tablecaption{The parameters of C$^{18}$O clumps}
\tabletypesize{\scriptsize}
\tablehead{
\colhead{Number} & \colhead{Source} & \colhead{$V_{\text{lsr}}$} & \colhead{$\Delta v$} & \colhead{$\theta_{\text{a}}$} & \colhead{$\theta_{\text{b}}$} & \colhead{$\theta_{\text{fwhm}}$} & \colhead{P.A.} & \colhead{Distance} & \colhead{$R_{\text{GC}}$} & \colhead{$T_{\text{ex}}$} & \colhead{log($N_{\text{H$_2$}}$)} & \colhead{$R_{\text{eff}}$} & \colhead{log($M_{\text{clump}}$)} & \colhead{log($M_{\text{vir}}$)} & \colhead{$\alpha_{\text{vir}}$} & \colhead{log($\Sigma$)} & \colhead{log($n_{\text{H$_2$}}$)} & \colhead{$\Delta v_{\text{th}}$} & \colhead{$\Delta v_{\text{nth}}$} & \colhead{Potential} \\
\colhead{} & \colhead{} & \colhead{(km s$^{-1}$)} & \colhead{(km s$^{-1}$)} & \colhead{($\arcsec$)} & \colhead{($\arcsec$)} & \colhead{($\arcsec$)} & \colhead{($\arcdeg$)} & \colhead{(kpc)} & \colhead{(kpc)} & \colhead{(K)} & \colhead{(cm$^{-2}$)} & \colhead{(pc)} & \colhead{($M_{\sun}$)} & \colhead{($M_{\sun}$)} & \colhead{} & \colhead{($M_{\sun}$ pc$^{-2}$)} & \colhead{(cm$^{-3}$)} & \colhead{(km s$^{-1}$)} & \colhead{(km s$^{-1}$)} & \colhead{}
\decimalcolnumbers
}
\startdata
1  & MWISP G24.672$-$0.152 & 112.34 & 2.05 & 65  & 145 & 83  & 111 & 6.69 & 3.48 & 14.0 & 21.7 & 1.3 & 3.1 & 3.1 & 0.9 & 2.3 & 3.8 & 0.15 & 2.04 &            \\
2  & MWISP G24.785$+$0.088 & 109.01 & 1.82 & 110 & 158 & 128 & 122 & 5.28 & 4.02 & 18.7 & 22.0 & 1.6 & 3.5 & 3.0 & 0.4 & 2.6 & 3.9 & 0.17 & 1.81 & \checkmark \\
3  & MWISP G24.475$+$0.479 & 102.84 & 1.38 & 81  & 150 & 101 & 87  & 5.65 & 3.81 & 13.8 & 21.7 & 1.3 & 3.1 & 2.7 & 0.4 & 2.3 & 3.8 & 0.15 & 1.37 &            \\
4  & MWISP G24.788$+$0.095 & 110.67 & 1.53 & 123 & 146 & 133 & 122 & 6.46 & 3.54 & 17.0 & 21.9 & 1.9 & 3.5 & 3.0 & 0.3 & 2.5 & 3.7 & 0.16 & 1.52 & \checkmark \\
5  & MWISP G24.491$-$0.043 & 110.50 & 1.97 & 94  & 105 & 99  & 43  & 6.68 & 3.46 & 10.2 & 21.4 & 1.4 & 2.8 & 3.1 & 1.9 & 2.0 & 3.4 & 0.12 & 1.97 &            \\
6  & MWISP G24.555$+$0.334 & 108.34 & 1.55 & 92  & 184 & 116 & 153 & 6.56 & 3.49 & 14.7 & 21.6 & 1.9 & 3.2 & 3.0 & 0.6 & 2.2 & 3.4 & 0.15 & 1.54 &            \\
7  & MWISP G24.540$-$0.283 & 96.83  & 1.30 & 85  & 85  & 85  & 81  & 5.54 & 3.87 & 12.1 & 21.4 & 0.9 & 2.5 & 2.5 & 1.0 & 2.1 & 3.7 & 0.14 & 1.29 &            \\
8  & MWISP G24.203$-$0.259 & 93.66  & 1.25 & 82  & 92  & 87  & 84  & 5.25 & 3.99 & 6.3  & 21.2 & 0.9 & 2.2 & 2.5 & 1.7 & 1.8 & 3.4 & 0.10 & 1.25 &            \\
9  & MWISP G24.457$+$0.191 & 119.17 & 1.75 & 83  & 150 & 103 & 35  & 7.23 & 3.38 & 10.6 & 21.5 & 1.7 & 3.1 & 3.0 & 0.8 & 2.1 & 3.4 & 0.13 & 1.75 &            \\
10 & MWISP G24.742$+$0.151 & 107.18 & 1.14 & 67  & 110 & 81  & 129 & 6.05 & 3.67 & 17.0 & 21.6 & 1.0 & 2.8 & 2.4 & 0.4 & 2.3 & 3.8 & 0.16 & 1.13 &            \\
\enddata
\tablecomments{Columns are same as Table \ref{appendix:the_parameters_of_12CO_clumps}.\\
(This table is available in its entirety in machine-readable form.)}
\label{appendix:the_parameters_of_C18O_clumps}
\end{deluxetable*}
\end{longrotatetable}

\clearpage
\bibliography{paper2}
\bibliographystyle{aasjournal}

\end{document}